\documentclass[camera,letterpaper,nomarginnotes,nonarrowgutter,hyperref]{jpaper}
\usepackage{amsmath,amssymb,amsfonts}
\usepackage{algorithm}
\usepackage[noend]{algpseudocode}
\usepackage{algorithmicx}
\usepackage{graphicx}
\usepackage{textcomp}
\usepackage{xcolor}

\usepackage{makecell}
\usepackage{array}
\usepackage{booktabs}
\usepackage[sort,compress]{cite}

\usepackage{xcolor, soul}
\usepackage{xspace}
\usepackage{multirow}
\usepackage{tabularx}
\usepackage{tikz}
\usepackage[bottom]{footmisc}

\usepackage[most]{tcolorbox}

\usepackage{fancyhdr}
\usepackage{mathtools}

\usepackage{subcaption}
\usepackage{booktabs}

\usepackage{pgfplots}
\pgfplotsset{compat=1.17} 

\usepackage{stfloats}

\usepackage{enumitem}

\usepackage{mwe}

\usepackage[most]{tcolorbox}

\definecolor{lavenderr}{rgb}{0.71, 0.49, 0.86}

\usepackage[most]{tcolorbox}   
\tcbset{
  observebox/.style={
    colframe=lavenderr!90!black,      
    colback=lavenderr!15,              
    boxrule=1pt,                   
    arc=1mm,                       
    left=0.1mm, right=0.1mm,       
    top=0.1mm, bottom=0.1mm,       
    fonttitle=\bfseries,           
    before skip=6pt, after skip=6pt
  }
}

\usepackage{graphicx}

\usepackage{tikzducks}

\usepackage[en-GB]{datetime2} 
\DTMsetup{
    showzone=false,   
    showseconds=false 
}

\usepackage{pifont}
\usepackage{path}
\definecolor{darkspringgreen}{rgb}{0.09, 0.45, 0.27}
\definecolor{denim}{rgb}{0.08, 0.38, 0.74}
\definecolor{darkolivegreen}{rgb}{0.33, 0.42, 0.18}
\definecolor{tangerine}{rgb}{0.95, 0.52, 0.0}
\definecolor{mahogany}{rgb}{0.75, 0.25, 0.0}
\definecolor{coolblack}{rgb}{0.0, 0.22, 0.44}
\definecolor{darkpink}{rgb}{0.91, 0.35, 0.6}
\definecolor{darkblue}{rgb}{0.0, 0.0, 0.67}
\definecolor{melon}{rgb}{0.97, 0.69, 0.67}

\definecolor{seagreen}{rgb}{0.18, 0.55, 0.34}
\definecolor{pred}{rgb}{0.7843, 0.0039, 0.3137} 

\definecolor{darkpink}{rgb}{0.88, 0.28, 0.54}
\definecolor{forestgreen}{rgb}{0.0, 0.27, 0.13}
\definecolor{amber}{rgb}{1.0, 0.49, 0.0}

\newcolumntype{Y}{>{\centering\arraybackslash}X}
\usepackage{tikz}

\newcommand{\squishlist}{
 \begin{list}{$\circ$}
  { \setlength{\itemsep}{0pt}
     \setlength{\parsep}{0pt}
     \setlength{\topsep}{3pt}
     \setlength{\partopsep}{0pt}
     \setlength{\leftmargin}{1em}
     \setlength{\labelwidth}{1em}
     \setlength{\labelsep}{0.5em} } }

\newcommand{\squishend}{
  \end{list}  }

\makeatletter
\g@addto@macro{\normalsize}{%
  \setlength{\abovedisplayskip}{4pt plus 0.5pt minus 1pt}
  \setlength{\belowdisplayskip}{4pt plus 0.5pt minus 1pt}
  \setlength{\abovedisplayshortskip}{0pt}
  \setlength{\belowdisplayshortskip}{0pt}
  \setlength{\intextsep}{3pt plus 1pt minus 1pt}
  \setlength{\textfloatsep}{7pt plus 1pt minus 1pt}
  \setlength{\skip\footins}{4pt plus 1pt minus 1pt}}
\makeatother

\usepackage{blindtext,graphicx}
\usepackage[absolute]{textpos}
\usepackage{datetime}

\definecolor{seagreen}{rgb}{0.18, 0.55, 0.34}
\definecolor{ballblue}{rgb}{0.13, 0.67, 0.8}

\definecolor{darkgreen}{rgb}{0.0, 0.44, 0.34}

\sloppypar

\definecolor{dollarbill}{rgb}{0.52, 0.73, 0.4}

\usetikzlibrary{patterns}
\usepackage[precision=2, unit=mm]{lengthconvert}

\usepackage[colorinlistoftodos,prependcaption,textsize=scriptsize]{todonotes}
\usepackage{marginnote} 
\definecolor{cyan(process)}{rgb}{0.0, 0.62, 0.82}

\definecolor{cadmiumgreen}{rgb}{0.0, 0.50, 0.29}

\newcommand\revref[1]{\hyperref[rev:#1]{#1}}

\definecolor{raspberry}{rgb}{0.89, 0.04, 0.36}

\definecolor{awesome}{rgb}{1.0, 0.13, 0.32}
\definecolor{cardinal}{rgb}{0.77, 0.12, 0.23}
\definecolor{cadet}{rgb}{0.33, 0.41, 0.47}
\definecolor{celadon}{rgb}{0.67, 0.88, 0.69}
\definecolor{persianblue}{rgb}{0.11, 0.22, 0.73}
\definecolor{ultramarine}{rgb}{0.07, 0.04, 0.56}
\definecolor{warmblack}{rgb}{0.0, 0.3, 0.3}

\definecolor{terracotta}{rgb}{0.89, 0.45, 0.36}

\definecolor{forestgreen(web)}{rgb}{0.13, 0.55, 0.13}

\definecolor{cardinal}{rgb}{0.77, 0.12, 0.23}
\definecolor{deeppink}{rgb}{1.0, 0.08, 0.58}
\definecolor{brightpink}{rgb}{1.0, 0.0, 0.5}
\definecolor{electricviolet}{rgb}{0.56, 0.0, 1.0}
\definecolor{brandeisblue}{rgb}{0.0, 0.44, 1.0}
\definecolor{carminered}{rgb}{1.0, 0.0, 0.22}

\usepackage{setspace}
\definecolor{acolor}{rgb}{0.0, 0.5, 1.0}
\definecolor{bcolor}{rgb}{0.54, 0.17, 0.89}
\definecolor{ccolor}{rgb}{0.4, 0.69, 0.2}
\definecolor{dcolor}{rgb}{0.92, 0.41, 0.12}
\definecolor{ecolor}{rgb}{0.6, 0.0, 0.156}
\definecolor{fcolor}{rgb}{0.106, 0.620, 0.467}

\definecolor{dogwoodrose}{rgb}{0.84, 0.09, 0.41}

\newcommand{\archname}{COSM}
\newcommand{\graycircled}[1]{%
  \mbox{\tikz[baseline=(char.base)]{
    \node[shape=circle, draw=black, fill=gray!30, inner sep=1pt, text=black, font=\scriptsize] (char) {#1};
  }}%
}
\newcommand{\circnum}[1]{\ding{\numexpr171+#1\relax}}

\sethlcolor{yellow!60}

\soulregister\ref7
\soulregister{\autoref}{1}
\soulregister\cite7
\soulregister{\citet}{1}
\soulregister{\citep}{1}
\soulregister{\eqref}{1}
\soulregister\figureautorefname7
\soulregister\sectionautorefname7
\soulregister\tableautorefname7
\soulregister\equationautorefname7
\soulregister{\graycircled}{1}

\newcommand{\minrevision}[1]{#1}
\newcommand{\revision}[1]{#1}
\newcommand\revisionlabel[1]{}

\renewcommand{\sectionautorefname}{Section}
\renewcommand{\figureautorefname}{Fig.}

\begin{document}
\bstctlcite{IEEEexample:BSTcontrol}

\title{\LARGE\archname{}: A Cooperative Scheduling Framework for \\ Concurrent PIM and CPU Execution on Mobile Devices}

\def\iscacameraready{} 
\newcommand{\hpcapubid}{0000--0000/00\$00.00}

\newcommand\iscaauthors{
Yilong Zhao$^{*1,2}$ \hspace{0.5em} Fangxin Liu$^{*1,2}$  \hspace{0.5em} Onur Mutlu$^3$ \hspace{0.5em} Mingyu Gao$^{4,2}$ \hspace{0.5em}
Jian Liu$^5$ \\ Haibing Guan$^{1}$ \hspace{0.5em} Li Jiang$^{\dagger 1,2,6}$ \\\\
}

\newcommand\iscaaffiliation{
\hspace{-2.5em} $^1$Shanghai Jiao Tong University \hspace{0.5em} $^2$Shanghai Qi Zhi Institute \hspace{0.5em} $^3$ETH Zurich \hspace{0.5em} $^4$ Tsinghua University \\ $^5$Beihang University  \hspace{0.5em} $^6$Huawei Technologies Co., Ltd.
}


\author{
\iscaauthors{}
\iscaaffiliation{}
}

\pagestyle{fancy}
\fancyhf{} 
\renewcommand{\headrulewidth}{0pt}


\renewcommand{\headrulewidth}{0pt}

\maketitle
\let\thefootnote\relax\footnotetext{$^*$ Yilong Zhao and Fangxin Liu contribute equally to this work. $^{\dagger}$ Li Jiang is the corresponding author.}

\newcommand{\iscaheight}{0mm}
\ifdefined\eaopen
\renewcommand{\iscaheight}{12mm}
\fi


\setcounter{page}{1}

\begin{abstract}
The development of on-device large language models (LLMs) is driven by the need for privacy and fast response times. Energy-intensive data transfer on mobile devices makes Processing-in-Memory (PIM) an effective solution. Due to stringent DRAM cost constraints, limited physical footprint on circuit boards, and the interaction between applications and LLMs, it is imperative for the CPU and PIM to operate concurrently within a shared memory space. However, challenges such as bank conflicts and bus congestion can arise, potentially diminishing the performance and energy benefits of PIM.

To address this challenge, we introduce \archname{}, a cooperative scheduling framework designed to facilitate the concurrent operation of PIM and CPU tasks on mobile platforms. Our key innovations include: 1) a low-interference PIM control interface that generates the maximum number of PIM commands without disrupting CPU memory accesses; 2) an idleness-aware scheduling method that integrates PIM commands into available idle time windows within the CPU's access sequence. \archname{} not only hides PIM execution latency from the CPU, but also overlaps PIM execution with data transfer. Experiments on concurrent execution of LLMs and mobile workloads, including mobile applications and compute-intensive kernels, demonstrate that \archname{} improves PIM throughput by up to 2.8\texttimes{} compared to the baseline scheduling method with less than 2.0\% CPU performance loss.
\end{abstract}


\section{Introduction}
\label{sec:introduction}

AI advancements are moving large language models (LLMs) from the cloud to edge devices~(e.g., mobile phones and PCs), enhancing privacy by keeping data local and enabling millisecond latency for interactive apps like voice assistants \cite{moshi_2024arXiv241000037D,minmo2025arXiv250106282C,shi2025voilavoicelanguagefoundationmodels}, real-time translation \cite{wang2025simultaneousmachinetranslationlarge,fu2025llmsachievehighqualitysimultaneous}, real-time video understanding \cite{lugaresi2019mediapipeframeworkbuildingperception}, and video editing \cite{tanveer2025motionbridgedynamicvideoinbetweening}. Industry trends show that companies such as Apple \cite{apple2024intelligence}, Huawei \cite{chen2025panguembeddedefficientdualsystem}, Qualcomm \cite{yao2024minicpm}, Samsung \cite{galaxyai2024}, and Vivo \cite{bluelmLu_2025_CVPR} have integrated AI on the device with models of 1B--3B parameters, highlighting ``localized intelligence'' as a key feature of future smart devices. Thus, efficiently running LLMs on resource-limited mobile devices has become a major challenge for both academia and industry.

Current on-device deployment strategies using Neural Processing Units (NPUs) have notable limitations. Despite high compute throughput, current devices' and NPUs' processor-centric architecture spends over 60\% of its total energy consumption on data movement during LLM inference, causing thermal challenges and shortening battery life \cite{qualcomm2024unlocking}. For instance, a 7B model on a mobile NPU can draw over 450~mA current \cite{vivo2024blueheart}. Moreover, the limited LPDDR5X memory bandwidth (typically $<$80~GB/s) also leads to latency fluctuations during long-context generation \cite{qualcomm2024unlocking}.

Processing-in-Memory (PIM) directly integrates computation units (PIM units) near memory banks in a DRAM chip to overcome limited memory bandwidth \cite{upmem8875680,AiMHW9731711,pushtap10.1145/3676642.3736120,UM-PIM10609641,PAPI10.1145/3676641.3716009,NeuPIMs10.1145/3620666.3651380,Unifying10.1007/978-981-95-1021-4_16,Pimba10.1145/3725843.3756121,attacc10.1145/3620665.3640422,pyramid10960667,Tesseract7284059,PEI7284077,google_nn_2021_9563028,sisa_10.1145/3466752.3480133,google_workload_10.1145/3173162.3173177,pian_10.1145/3676641.3716267,primer_Mutlu2023,enabling_8405955,benchmarking_9771457,evaluating_10158216,benchmarking2_9651614}.
This innovation enhances internal bandwidth (i.e., the bandwidth from the memory banks and row buffers to either I/O circuitry or PIM units) by processing data locally, thereby bypassing the bottleneck of external bandwidth (i.e., the bandwidth across the memory bus).
Samsung's LPDDR5-PIM prototype showed a 70\% reduction in power consumption and 102.4~GB/s memory bandwidth in mobile settings \cite{samsunglpddr5pim10254711}.
Thus, PIM is emerging as a promising and effective architectural paradigm to overcome the energy and memory bottlenecks of mobile AI hardware \cite{samsunglpddr5pim10254711}.

Some DRAM-PIM designs, like UPMEM's DDR4-PIM \cite{upmem8875680}, enforce static partitioning between CPU and PIM memory spaces, which incurs substantial memory reservation overhead for LLMs and significant data movement between CPU and PIM units.
Shared-memory PIM architectures, where the CPU and PIM units physically share the same memory space rather than operating in isolated memory spaces, address this problem by using OS-managed logical isolation instead but they introduce new challenges in memory management~\cite{UM-PIM10609641,PIM-MMU10764703} and bandwidth scheduling~\cite{AsyncDIMM10946818,Chopim9138972,F3FS11096393,ComPASS10.1145/3725843.3756017}.
 While recent works aim to tackle these issues, the industry has not yet deeply explored this direction. We observe complementary bandwidth usage: \emph{CPUs use high external DRAM bandwidth but low internal DRAM bandwidth, opposite to PIM units.} This allows us to leverage idle internal bandwidth of CPU workloads for PIM tasks, enhancing memory efficiency and offering dual benefits for mobile devices: reduced overhead due to dynamic sharings of banks between CPU and PIM units and lower power consumption due to less data movement.

The shared memory design aims to harvest idle DRAM bandwidth for PIM workloads without significantly impacting CPU performance. Current PIM scheduling frameworks are hindered by memory scheduling and PIM control interface constraints. Memory scheduling strategies have trade-offs: CPU-first scheduling \cite{Chopim9138972} maintains CPU latency but underutilizes internal bandwidth, while row-hit-aware scheduling \cite{F3FS11096393,AsyncDIMM10946818} optimizes bandwidth utilization and PIM performance but degrades CPU performance. Both strategies struggle to balance PIM and CPU performance. Additionally, PIM control limitations further exacerbate the scheduling difficulty. PIM architectures with fine-grained commands minimize CPU blocking, but can saturate the command bus, reducing PIM throughput \cite{HBM-PIM-HW9365862,AiMHW9731711}. In addition, CPU-initiated data transfers in PIM workloads for staging inputs and collecting results, referred to as CPU-mediated transfer, share data paths with CPU access, causing contention and degrading CPU performance.

In this work, we present \archname{}, a novel \uline{co}operative \uline{s}cheduling framework for concurrent PI\uline{M}/CPU execution designed to balance CPU and PIM performance trade-offs. 
We first propose a \emph{low-interference PIM control interface} that includes two key mechanisms: (1) a \emph{preemptable PIM execution command} to mitigate command bus contention while ensuring responsive CPU access, and (2) a \emph{bandwidth-decoupled data transfer command} to prevent PIM data transfer from stalling CPU accesses. 
Built on this interface, we propose an \emph{idleness-aware scheduling strategy} in the memory controller.
Specifically, by monitoring the CPU access queue, the controller identifies idle time windows on both the memory bus and banks.
It then schedules PIM commands, including PIM execution and data transfer, within these idle time windows, to fully exploit idle bandwidth while minimizing interference with CPU memory accesses.
This strategy enhances resource utilization while incurring minimal impact on CPU performance.

We make the following contributions:
\begin{itemize}
    \item We provide key observations that identify critical factors affecting the performance of both CPU and PIM workloads, offering design insights for PIM systems. While CPU workloads are sensitive to memory access latency, existing PIM designs cause significant interference during both PIM execution and CPU-mediated data transfers. This is a conflict rooted in both the PIM control interface and memory scheduling strategy. Particularly, CPU-mediated data transfers within PIM workloads incur substantial performance degradation for CPU tasks.
    \item We propose \archname{}, a new cooperative scheduling framework for concurrent PIM/CPU execution on mobile devices.
    At the interface level, \archname{} introduces a low-interference PIM control interface tailored to our observations of CPU and PIM workload conflicts. Building on this, our scheduling policy precisely coordinates PIM command dispatch to exploit CPU idle time windows of CPU access, keeping CPU access latency low while maximizing PIM performance.
    \item Our comprehensive experiments show that \archname{} improves LLM throughput on PIM by 2.6\texttimes{} with less than 2.2\% performance degradation on concurrent CPU workloads. 
\end{itemize}

\section{Background}
\label{sec:background}

\subsection{Hierarchical Architecture of DRAM}
\label{sec:background:DRAM}

In modern DRAM, the latency of control commands (e.g., opening a row in LPDDR5 \cite{JEDECLPDDR52023} requires $tRP + tRCD \approx 30$ DRAM clock cycles) is much longer than the duration of the burst length ($tBL = 8$ DRAM clock cycles), where $tRP$, $tRCD$, and $tBL$ refer to row precharge time, the row address strobe to column address strobe delay, and the burst length \cite{salp_6237032,tl_dram_6522354}.
To mitigate this overhead, DRAM adopts a hierarchical architecture. Multiple banks in a rank can operate independently, and row operations can overlap with each other across banks that share a common memory bus. 

\begin{figure}[h]
    \centering
    \includegraphics[width=0.9\linewidth]{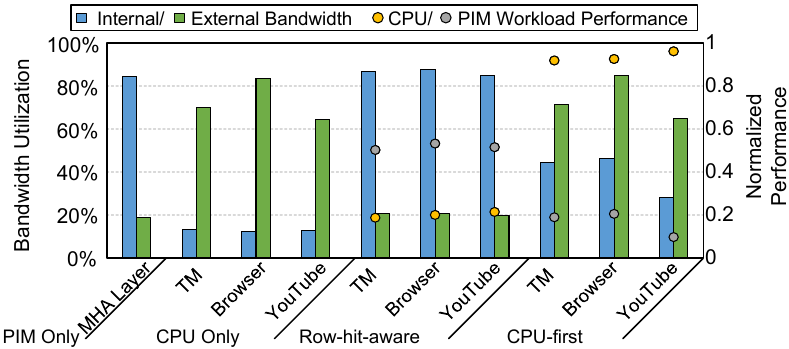}
    \caption{Internal/external DRAM bandwidth utilization of CPU and PIM workloads, and CPU/PIM workload performance under different scheduling strategies when concurrently executing on physically shared memory space. (Note: TM stands for TencentMeeting.)}
    \label{fig:ob-inexternal-bandwidth}
\end{figure}

As DRAM density increases with increasing bank counts, the overhead of row opening emerges as a fundamental scaling bottleneck. Although more banks offer higher theoretical parallelism, fixed command bandwidth severely restricts maximum internal bandwidth utilization. For example, in a typical 2-rank LPDDR5 per-channel mobile phone setup, each rank has 32 banks~\cite{samsunglpddr5pim10254711} ($\#bank=32$). For a workload with a row hit rate $R_h$, the upper bound under saturated external bandwidth (i.e., assuming 100\% external bandwidth utilization) is:
\begin{equation}
    Util_{i} = \dfrac{tBL + (tRP + tRCD) \cdot (1 - R_h)}{\#bank \cdot tBL}
\end{equation}
The $\#bank$ term in the denominator illustrates that the shared command bus serializes bank accesses, causing the denominator to far outweigh the numerator.
This reveals a fundamental problem: \emph{even with completely random access that leads to zero hit rate ($R_h=0$), the utilization of internal bandwidth cannot exceed $15\%$.}
Real-world workloads typically achieve far lower utilization, as demonstrated in \figureautorefname{}~\ref{fig:ob-inexternal-bandwidth}.

\subsection{DRAM-based PIM}
\label{sec:background:PIM}

Unlike traditional heterogeneous systems (e.g., CPU-GPU systems) where all compute units share a unified DRAM controller interface, PIM units are spatially distributed across DRAM banks and access memory through dedicated intra-bank pathways.
This fundamental architectural distinction requires specialized interfaces for PIM systems. 
Current implementations predominantly adopt two types of PIM interfaces:

\subsubsection{Two-Host Design} 

A PIM unit functions independently from the CPU, with its own instruction sequencer and local \minrevision{DRAM access~\cite{upmem8875680,ortega2024pimainovelarchitecturehighefficiency,ALPHA-PIM-11242082,benchmarking_9771457,evaluating_10158216,pygim10.1145/3700434,sparsep_9912078}}. During PIM operations, CPU access to these banks is blocked to prevent DRAM state corruption. Completion of PIM tasks is detected through polling, which checks status registers. After execution, the memory controller must resynchronize DRAM for CPU access, causing significant switching overhead between CPU and PIM access.

\subsubsection{Single-Host Design}

PIM units use extended DRAM commands for precise control, optimized for specific PIM workloads due to limited \minrevision{command encoding space \cite{AiMHW9731711,AiMSW9895629,HBM-PIM-HW9365862,HBM-PIM-SW9499894,AsyncDIMM10946818,Chopim9138972,pian_10.1145/3676641.3716267}}. Recent advances include translation tables that map high-level operations to these commands, enhancing flexibility. This design, unlike the two-host model, keeps the memory controller fully aware of changes in DRAM state during PIM operations. Its centralized scheduling allows fine-grained interleaving of CPU and PIM commands without extra synchronization, supporting efficient concurrency by removing conservative timing or polling requirements.

\section{Key Observations and Design Implications}
\label{sec:motivation}

We analyze CPU and PIM workload interactions in a shared-memory CPU/PIM hybrid system with concurrent execution. 
We present three key observations affecting performance and analyze concurrent execution techniques. These insights drive our interface and scheduling co-design in \sectionautorefname{}~\ref{sec:interface} and \ref{sec:scheduling}.

\subsection{Effect of Latency Interference on Memory Access}
\label{sec:motivation:cpu-perf-with-latency}

\begin{figure}[h]
    \centering
    \includegraphics[width=0.9\linewidth]{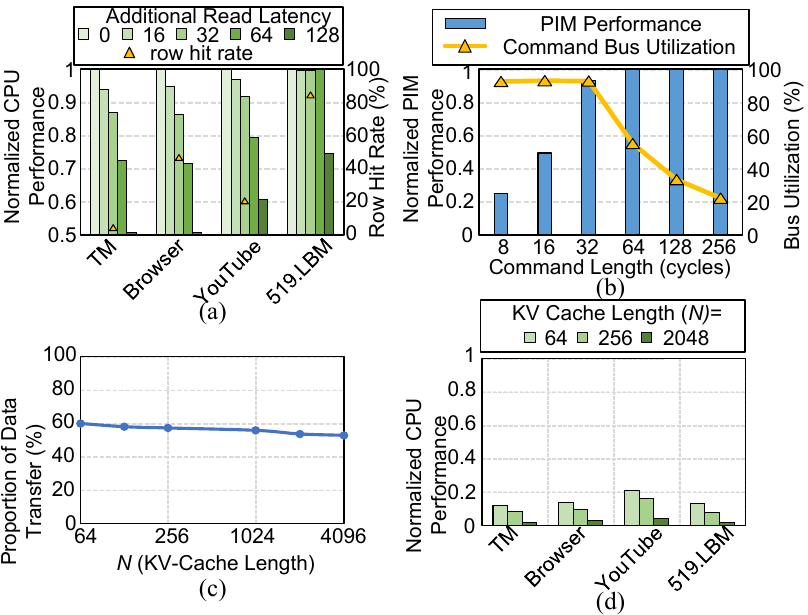}
    \caption{(a) CPU workload performance under injected read latency. (b) PIM workload performance and command bus occupation across command lengths. \revision{Performance is normalized to the peak performance under an unsaturated command bus (command length $>=64$)}. (c) Proportion of CPU-mediated transfer in an attention layer inference of DeepSeek-R1-1.5B. \minrevision{(d) CPU workload performance when concurrently executing with CPU-mediated PIM data transfer.}}
    \label{fig:ob-cpu-lat-and-pim-len}
\end{figure}

To characterize the impact of memory-side contention, we conduct a sensitivity study on CPU performance in response to increased memory latency.
We simulate PIM-induced access delays by injecting additional CPU read latency on a CPU-only system across three real-world applications and a SPEC 2017 benchmark. As shown in Figure \ref{fig:ob-cpu-lat-and-pim-len}(a), a 16-cycle latency increase reduces CPU performance by more than 5\%. 
As latency increases, performance significantly degrades: e.g., a 128-cycle latency reduces CPU performance by more than 40\% for some workloads. This confirms that CPU workload performance is sensitive to memory access latency.

The results of this study offer two key guidelines for the formulation of memory scheduling strategies in the concurrent execution of CPU-PIM hybrid systems. 
First, the system must enable fast preemption of PIM operations to reduce CPU request latency. This requires interrupting PIM operations and minimizing the switching overhead between CPU and PIM commands. A single-host design with fine-grained PIM command control is preferred, as a two-host design introduces significant switching latency, and coarse-grained PIM commands hinder timely CPU access. 
Second, memory scheduling should give precedence to CPU memory accesses and restrict PIM operations to periods when the memory is idle, thereby guaranteeing minimal disruption. Collectively, these guidelines require a closely integrated design of the PIM execution interface alongside the memory scheduler.

\subsection{Effect of PIM Command Granularity on Performance}
\label{sec:motivation:pim-perf-with-length}

We study how the granularity of PIM execution commands affects PIM performance, defining the ``\emph{command length}'' as the number of cycles PIM units can execute autonomously per command. Figure \ref{fig:ob-cpu-lat-and-pim-len}(b) shows that longer command lengths improve PIM performance in a single-host system with 2 LPDDR5 ranks per channel and 32 banks in total. A command with a length of $\geq$ 128 keeps the command bandwidth below 40\%, while a length of $\geq$ 64 is needed for full bank-level parallelism across 32 banks~(taking into account the opening overheads of rows). Shorter command lengths cause command bus congestion and external DRAM bandwidth underutilization.

\emph{This finding is in tension with the observation in \sectionautorefname{}~\ref{sec:motivation:cpu-perf-with-latency}: longer command lengths boost PIM performance, but shorter ones are crucial to minimize CPU memory latency.} To eliminate this tension in existing fixed-length command architectures, we propose \emph{preemptable PIM execution commands} in the \archname{} framework to balance extended command benefits with CPU access needs.

\subsection{Interference from CPU-mediated PIM Data Transfer}
\label{sec:motivation:external}

Although PIM workloads primarily leverage internal DRAM bandwidth between PIM units and their local banks, the host CPU still needs to transfer input data to PIM units and collect results from PIM units.
These necessary CPU-mediated transfers create memory contention for co-executing CPU workloads that share the same DRAM devices.

To quantify the impact of this bandwidth contention, we first measure the proportion of CPU-mediated transfers during an attention layer inference of a DeepSeek-R1-1.5B model \cite{deepseekai2025deepseekr1incentivizingreasoningcapability} in the decoding phase on the same system as \sectionautorefname{}~\ref{sec:motivation:pim-perf-with-length}, as shown in \figureautorefname{}~\ref{fig:ob-cpu-lat-and-pim-len}(c). 
The command length of each PIM execution is set to 128 cycles.
When the sequence length dimension $N$ of Key-Value (KV) cache is 64, CPU-mediated transfers account for over 60\% of the overall inference time. 
As $N$ increases during inference, this proportion gradually decreases to around 50\% when $N$=4k, demonstrating that CPU-mediated transfers still constitute a substantial fraction of the overall execution time.

\figureautorefname{}~\ref{fig:ob-cpu-lat-and-pim-len}(d) further demonstrates the interference caused by CPU-mediated transfers on concurrent CPU workloads. 
Both CPU-mediated transfers of PIM workload and conventional CPU memory requests in CPU workload share the same memory request queue, which is scheduled using the classic First-Ready, First-Come First-Serve (FR-FCFS) policy \cite{FR-FCFS10.1145/339647.339668,zuravleff1997controller}.
We observe that CPU workload performance degrades by more than 80\% when concurrent with CPU-mediated transfers.
This severe slowdown arises because CPU-mediated transfers exhibit bursty access patterns with high row-buffer locality. 
Under the FR-FCFS scheduler, such bursts monopolize the memory controller (as described in \cite{memory_performance_attack_268461,fairscheduling4408252,parallelism-aware4556716}) by repeatedly hitting open rows, thereby starving CPU workloads, whose requests often require costly row conflicts.

\revision{\emph{These results reveal a critical insight: the effect of the observed 80\% CPU performance degradation during the 50\% data transfer can translate into a 40\% reduction in overall CPU performance.}}
This disproportionate impact shows that CPU-mediated transfers must be treated as first-class citizens in memory scheduling in concurrent CPU and PIM access scenarios.
Consequently, CPU-mediated transfers must be governed by the same scheduling discipline as PIM units' execution, i.e., scheduled during idle time windows.
However, implementing the CPU-mediated transfers requires simultaneous availability of \emph{both external and internal} idle time windows, making it difficult to fully utilize memory idle time windows, limiting PIM throughput. 
\revision{To overcome this, \archname{} introduces \emph{bandwidth-decoupled CPU-mediated transfer commands} that decouple the usage of external and internal bandwidth, eliminating the need for simultaneous external and internal idle time window availability.}

\subsection{Concurrent CPU-PIM Execution Methods}
\label{sec:motivation:scheduling}

\begin{figure}[h]
    \centering
    \includegraphics[width=0.85\linewidth]{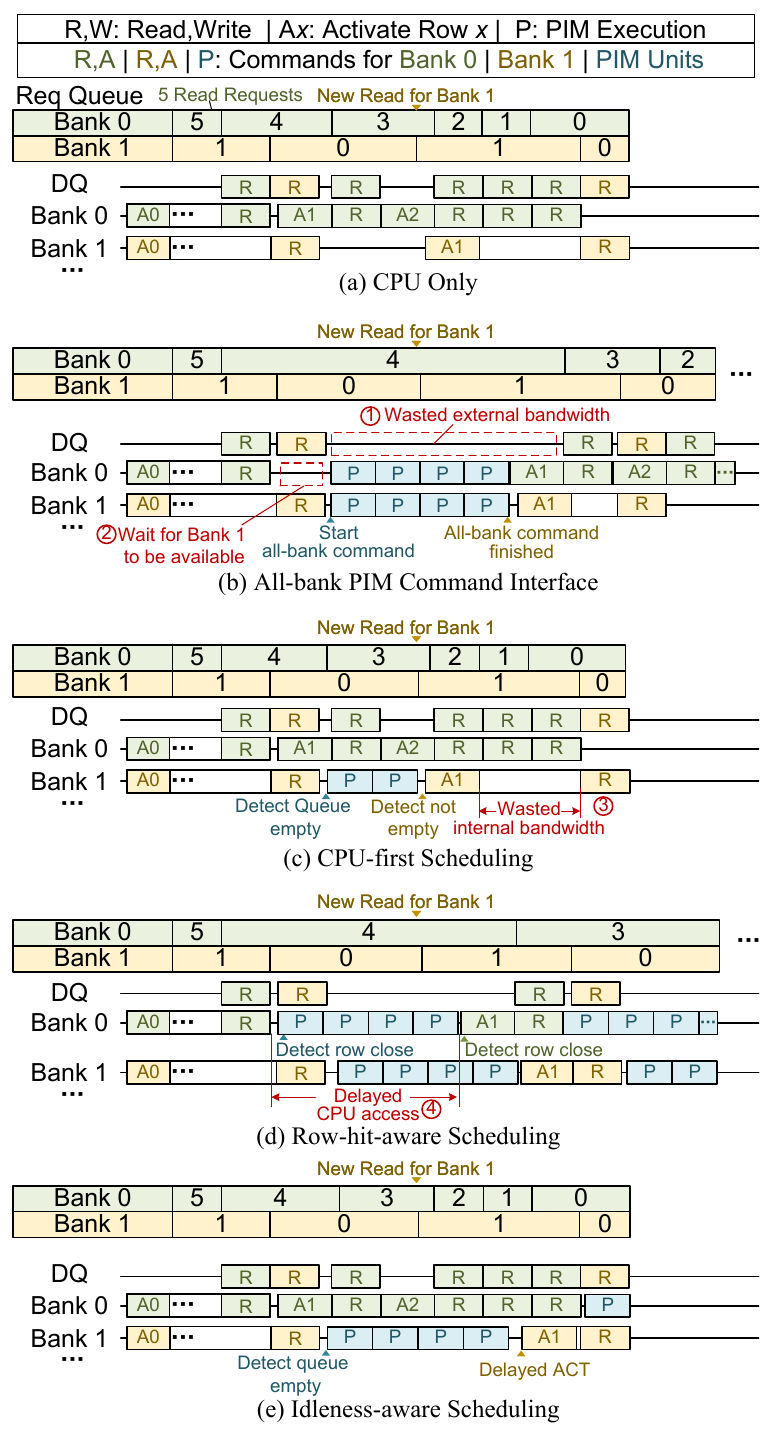}
    \caption{(a) An example of FR-FCFS scheduling for a CPU-only workload and (b-e) methods for concurrent PIM/CPU execution. To simplify the diagram, precharge is combined with activation, and row activations prior to PIM executions are not shown.}
    \label{fig:scheduling-methods}
\end{figure}

Based on the observations above, we analyze several techniques for concurrent execution of PIM and CPU on shared memory banks. 
\figureautorefname{}~\ref{fig:scheduling-methods}(a) depicts CPU memory request processing under FR-FCFS, showing memory bus and bank occupancy along with per-bank queue lengths.
\figureautorefname{}~\ref{fig:scheduling-methods}(b)-(d) compare three concurrency techniques under the same CPU workload.
Each PIM workload consists of one burst write of CPU-mediated transfer (co-scheduled with normal CPU access under FR-FCFS) followed by multiple execution commands.

\subsubsection{All-bank PIM Command Interface}
The \emph{all-bank} command mitigates command bandwidth bottlenecks by invoking all PIM units with one PIM command \cite{HBM-PIM-HW9365862}. As shown in \figureautorefname{}~\ref{fig:scheduling-methods} (b), this method integrates a PIM command amidst CPU memory requests. However, it causes bandwidth underutilization: all-bank command execution leaves no bank available for CPU accesses, making external bandwidth idle (\circnum{1}), and PIM commands must wait for all banks to become available, causing latency (\circnum{2}).

\subsubsection{CPU-first Scheduling}
Chopim \cite{Chopim9138972} introduces a memory scheduling algorithm based on the idea that CPU workloads are sensitive to memory latency. It blocks PIM commands once the memory queue is detected as not empty, as shown in \figureautorefname{}~\ref{fig:scheduling-methods}(c). While PIM commands minimally impact CPU latency, bandwidth utilization remains suboptimal. 
When a request targets Row 1 of Bank 1, the scheduler immediately stops issuing PIM commands. However, due to bus contention, there is a latency gap between row activation command (\texttt{ACT}) and data access, leading to wasted internal bandwidth (\circnum{3}).

\subsubsection{Row-hit-aware Scheduling}
AsyncDIMM~\cite{AsyncDIMM10946818} and F3FS~\cite{F3FS11096393} suggest that the memory controller switches between the PIM command queue and the CPU request queue when a row-close command (\texttt{PRE}) is executed, benefiting memory-intensive tasks with high row hit rates by balancing the CPU and PIM scheduling. However, with this policy, CPU performance suffers in compute-intensive tasks with random access. As shown in \figureautorefname{}~\ref{fig:scheduling-methods}(d), closing row 1 in bank 0 forces the memory controller to switch to the PIM command queue, causing CPU access delays and higher memory latency (\circnum{4}).

As such, prior scheduling methods cannot both provide low CPU memory latency and allow PIM workloads to fully utilize the temporary idle periods across banks on mobile devices.
Furthermore, prior scheduling methods lack a dedicated scheduling policy for CPU-mediated data transfers, which interferes with CPU workloads as observed in \sectionautorefname{}~\ref{sec:motivation:external}.  
\section{Architecture Overview}
\label{sec:architecture}

We introduce a new scheduling framework in the memory controller to enhance the use of PIM workload bandwidth without affecting CPU memory latency. Our approach opportunistically schedules PIM operations during CPU request idle times, ensuring minimal CPU interference. \archname{} employs a dual-path SW/HW co-design. First, we extend the DRAM command set to enable a low-interference PIM control interface~(see \sectionautorefname{}~\ref{sec:interface}), forcing PIM tasks to pause for high-priority CPU requests, and create dedicated commands for CPU-mediated data transfers that decouple internal and external bandwidth usage. Second, we develop an idleness-aware scheduling policy~(see \sectionautorefname{}~\ref{sec:scheduling}) that inserts PIM commands during idle time windows of CPU access while maintaining CPU latency guarantees, by enhancing the memory controller hardware.

\begin{figure}[h]
    \centering
    \includegraphics[width=0.9\linewidth]{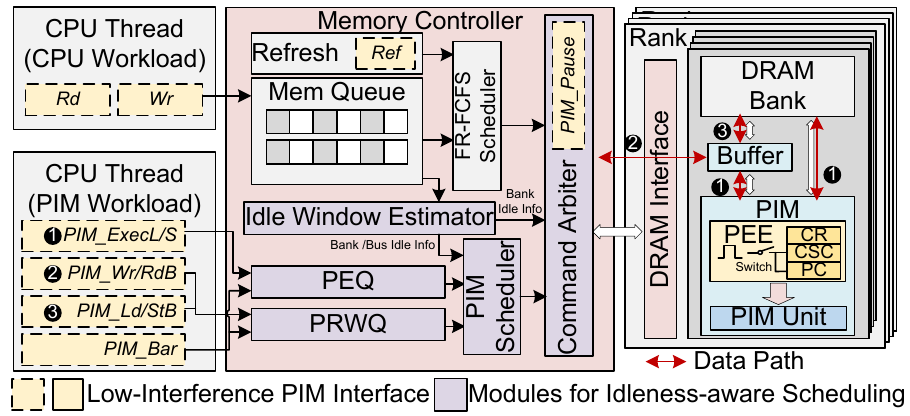}
    \caption{\archname{}'s memory controller architecture and memory interface}
    \label{fig:architecture}
\end{figure}

Figure~\ref{fig:architecture} shows the architecture of the \archname{} memory controller. We extend the software layer on the CPU side to accommodate a low-interference memory interface. 
A PIM Execution Engine (PEE) in PIM units manages preemptable command execution and prevents command bus saturation.
\revision{We use an SRAM buffer next to each DRAM bank to buffer both the operands for PIM execution and the data staged by the decoupled-bandwidth CPU-mediated transfer commands.}
facilitate both CPU-mediated transfers and PIM execution. Within the memory controller, we introduce two dedicated PIM queues: the PIM Execution Queue (PEQ), which buffers commands for PIM computations (e.g., \texttt{PIM\_Exec(L)}, \texttt{PIM\_Exec(S)}), and the PIM Read/Write Queue (PRWQ), which manages CPU-mediated transfer commands (e.g., \texttt{PIM\_Ld/StB}, \texttt{PIM\_Rd/WrB}). The PIM scheduler is responsible for selecting candidate commands from these queues, while the idle time window estimator (IWE) analyzes patterns in the CPU access queue to predict forthcoming bank- and bus-level idle time windows in between CPU accesses. The Command Arbiter consolidates and chooses from three command sources: 1) the CPU memory access candidate from the traditional FR-FCFS scheduler, 2) the PIM candidate from the PIM scheduler, and 3) pause commands to pause PIM execution (\texttt{PIM\_Pause}) generated by the Command Arbiter itself. \revision{The Command Arbiter employs a strict priority policy to decide the final sequence for issuing DRAM commands (\sectionautorefname{}~\ref{sec:scheduling:scheduler}).}

The controller operates in a four-stage pipeline. First, requests are queued by command type. Second, the IWE examines access queues to predict idle time windows, sending its predictions to the Command Arbiter and PIM scheduler. Third, the FR-FCFS and PIM schedulers each select commands based on their prioritization policies. 
\revision{Finally, the Command Arbiter makes decision: If a PIM-induced stall is predicted for a CPU memory request, the Command Arbiter issues a \texttt{PIM\_Pause} to prevent delaying the CPU accesses. Otherwise, priority is granted to the FR-FCFS scheduler (i.e., serving CPU access and refresh requests). If no such requests are avaliable, a PIM command is scheduled.}

\section{Low-Interference PIM Control Interface}
\label{sec:interface}

\archname{} introduces a refined PIM control interface that explicitly provide two seperate new features: (1) preemptable PIM execution for compute phases, and (2) bandwidth-decoupled commands for CPU-mediated data movement. This enables fine-grained scheduling control, allowing PIM operations to yield instantly to CPU requests and exploit fragmented idle time windows for PIM commands without compromising data staging efficiency.

\subsection{Preemptable PIM Execution commands}
\label{sec:interface:pausable}

The proposed preemptable PIM control interface extends standard DRAM commands with two critical additions: PIM execution command family (\texttt{PIM\_Exec}) that allow PIM units to automatically execute computations continuously for an extended period without requiring additional trigger commands, and a PIM pause command (\texttt{PIM\_Pause}) that enables preemption by halting PIM execution. 
Compared to the fixed-length command design discussed in \sectionautorefname{}s~\ref{sec:motivation:cpu-perf-with-latency} and \ref{sec:motivation:pim-perf-with-length}, the preemptable commands enable immediate reaction to incoming CPU access while avoiding command bus saturation that would otherwise degrade PIM performance. 
The scope of all the PIM command are at the bank level. 
We employ a hardware-command co-design approach to achieve preemptable computation. Managed by the memory controller, this interface allows fine-grained execution of PIM workloads while enabling low CPU memory access latency.

\begin{figure}[h]
    \centering
    \includegraphics[width=\linewidth]{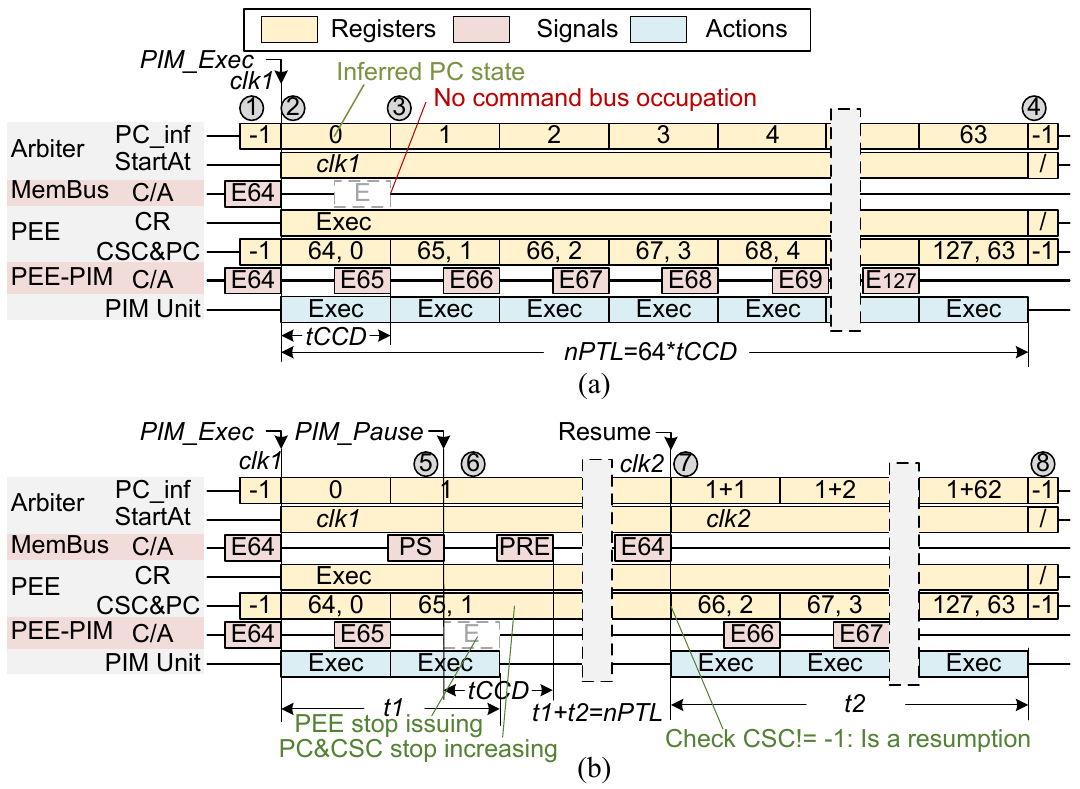}
    \caption{Timing diagram of PIM unit, memory bus, Command Arbiter, and PEE register states of preemptable PIM execution command when (a) no CPU memory access is present and (b) \texttt{PIM\_Pause} is sent by Command Arbiter when processing the second column. E$n$: PIM execution on column $n$, PS: PIM pause command.}
    \label{fig:pausable-intf}
\end{figure}

\textbf{PIM Execution Commands.}
In our setup, the execution commands \texttt{PIM\_Exec} for PIM are tailored for LLM operations such as MAC and softmax. 
\figureautorefname~\ref{fig:pausable-intf}~(a) illustrates the execution of a PIM command and the register states of the PEE. 
\revision{
If a command with start column 64 is issued at $clk1$, the PEE module of the PIM unit switches from state \graycircled{1} to \graycircled{2} by setting the Column State Counter (CSC) with the start address, recording the command type in the Command Register (CR), and setting the PIM Counter (PC) as 0. For every $tCCD$~(column-to-column delay), the PEE autonomously increments the Column State Counter and PIM Counter and issues commands in Command Register using the Column State Counter as the column address (\graycircled{3}). 
The Command Arbiter can always synchronize the PIM Counter state by inferring the PIM Counter : PC\_inf=$(clk-clk1)/tCCD$, where $clk$ is the current clock cycle.
This procedure halts after $nPTL$ cycles~(i.e., PIM execution command length predefined by the DRAM configuration register) by comparing PIM Counter value with the $nPTL/tCCD$ value if there is no CPU memory access, and PEE returns to its default state (\graycircled{4}). 
}
This reduces the usage of the command bus by $nPTL/tCCD$ times compared to fine-grained commands (see \sectionautorefname{}~\ref{sec:motivation:pim-perf-with-length}). 
In the setup shown in \figureautorefname~\ref{fig:ob-cpu-lat-and-pim-len}~(b) (2 ranks \texttimes 16 banks), $nPTL$ needs to be at least 64 cycles for more flexibility, with lower values leading to command bus contention.

\textbf{PIM Pause Command.}
The \texttt{PIM\_Pause} command ensures timely CPU request handling by pausing PIM execution. 
\figureautorefname~\ref{fig:pausable-intf}~(b) shows a CPU memory request interrupting PIM execution when the CPU reads from a bank still processing column 65 at time $clk2=clk1+tCCD+1$.
The memory controller's Command Arbiter sends \texttt{PIM\_Pause} to the bank. Upon receiving it, the PEE completes the current column operation in up to $tCCD$ cycles, freezes the Column State Counter (by opening the Switch of PEE in \figureautorefname{}~\ref{fig:architecture}), and releases the bus control (from \graycircled{5} to \graycircled{6}). The Command Arbiter can directly infer the PIM Counter state by calculating the interval between $clk1$ and the end of the current execution ($t1=2tCCD$). After CPU access, the Command Arbiter reissues the origin PIM execution command, resuming from the Column State Counter-stored column address (\graycircled{7}). Knowing that columns 64-65 are completed, the Command Arbiter determines the remaining execution time as $t2=nPTL-2tCCD=62tCCD$ (\graycircled{8}), synchronizing with the PIM Counter without extra signaling and minimizing timing constraints.

\textbf{Timing Constraints.} 
\tableautorefname{}~\ref{tab:timing_constrain} provides the timing constraints for preemptable PIM commands. Similar to DRAM read/write commands, a PIM execution command can only be issued after $tRCD$ after row activation. The \texttt{PIM\_Pause} follows at least $tCCD$ cycles after PIM execution command to ensure PIM unit column access completion. Post \texttt{PIM\_Pause}, if the execution command is load-only (\texttt{PIM\_Exec(Ld)}), a \texttt{PRE} can be issued after $tRTP$ cycles. For execution commands with data stores (\texttt{PIM\_Exec(St)}), \texttt{PRE} must wait for data stabilization, requiring $tWR$. This longer latency necessitates careful design to maximize intermediate result reuse and minimize store operations.

\begin{table}[h]
    \centering
    \caption{\revision{Timing Constraint of PIM Commands}}
    \label{tab:timing_constrain}
    \resizebox{\linewidth}{!}{
    \setlength{\tabcolsep}{4pt} 
    \begin{tabular}{c|c|c|c|c}
        \hline
        Scope   & Previous  & Next          & Min. delay        & Conflict  \\ \hline\hline
        \multirow{4}{*}[-15pt]{Bank}
                & \texttt{ACT}       & \begin{tabular}{c} \texttt{PIM\_Exec} \\ \texttt{PIM\_LdBuf/StBuf} \end{tabular}   
                                            & tRCD              & \multirow{3}{*}[-10pt]{\begin{tabular}{c} DRAM \\ Array \end{tabular}}
                                                                            \\ \cline{2-4}
                & \begin{tabular}{c} \texttt{PIM\_Exec} \\ \texttt{PIM\_Ld/StBuf} \end{tabular}   
                            & \texttt{PIM\_Pause}    & $tCCD$              &           \\ \cline{2-4}
                & \begin{tabular}{c}\texttt{PIM\_Pause}\\(for \texttt{PIM\_Exec(Ld)})\end{tabular}   
                            & \texttt{PRE}           & $tRTP$              &           \\ \cline{2-4}
                & \begin{tabular}{c}\texttt{PIM\_Pause} \\ (for \texttt{PIM\_Exec(St)})  \end{tabular}
                            & \texttt{PRE}           & $tCCD$+$tWR $         &           \\ \cline{2-5}
                & \begin{tabular}{c} \texttt{PIM\_RdBuf/WrBuf} \\ \texttt{PIM\_LdBuf/StBuf} \\  \end{tabular}
                            & \begin{tabular}{c} \texttt{PIM\_RdBuf/WrBuf} \\ \texttt{PIM\_LdBuf/StBuf} \end{tabular}  
                                            & $tBL$               & \begin{tabular}{c} PIM \\ Buffer \end{tabular} 
                                                                            \\ \hline
        Channel & \begin{tabular}{c} \texttt{PIM\_RdBuf/WrBuf} \\ \texttt{Read/Write} \\  \end{tabular}
                            & \begin{tabular}{c} \texttt{PIM\_RdBuf/WrBuf} \\ \texttt{Read/Write} \end{tabular}  
                                            & $tBL$               & \begin{tabular}{c} Memory \\ Bus \end{tabular} 
                                                                            \\ \hline\hline
    \end{tabular}
    }
\end{table}

\textbf{Semantic Guarantees of \texttt{PIM\_Pause}.} 
The correctness of the PIM pause mechanism relies on strict adherence to DRAM timing constraints and three fundamental properties. 
First, it guarantees \textit{column atomicity} by pausing only at column boundaries (after integer multiples of $tCCD$), ensuring deterministic progress tracking for both the PIM unit and the memory controller.
Second, it preserves all intermediate data in PIM buffers and architectural states during suspension, preventing context corruption. 
Third, it correctly restores PIM state by reactivating the original row and reissuing the paused command; the PIM unit then leverages frozen PIM Counter to resume execution precisely at the pause point.
Collectively, these guarantees ensure that preempted CPU accesses and DRAM refreshes do not compromise the correctness of PIM execution.

\subsection{Bandwidth-Decoupled CPU-mediated PIM Data Transfer}
\label{sec:interface:decoupled}

Conventional PIM data transfers are based on standard DRAM read/write commands, which could require a long  sequence of row activation, column access, and external data transfer. Moreover, the CPU-mediated transfer must occur continuously, locking the channel for the entire duration of the tranfer.

\begin{figure}[h]
    \centering
    \includegraphics[width=0.9\linewidth]{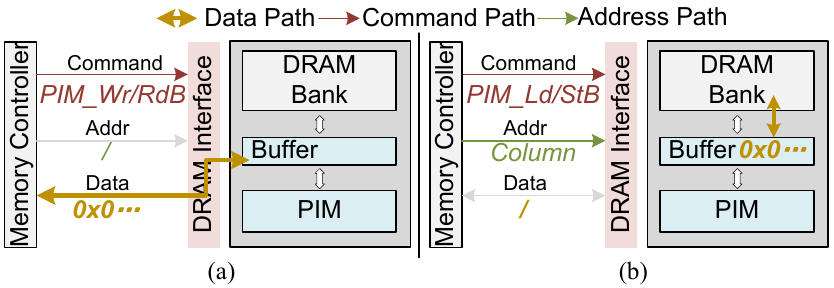}
    \caption{The data, command, and address path of (a) \texttt{PIM\_WrBuf/RdBuf} and (b) \texttt{PIM\_LdBuf/StBuf}}
    \label{fig:decoupled}
\end{figure}

We introduce a mechanism that decouples the usage of external bandwidth from that of internal bandwidth during CPU-mediated data transfer. As depicted in \figureautorefname{}~\ref{fig:decoupled}, \revision{the memory interface employs two-phase reads and writes via four commands: \texttt{PIM\_RdBuf} and \texttt{PIM\_WrBuf} for \emph{external transfers} between the memory controller and the PIM buffer through the memory bus, and \texttt{PIM\_LdBuf} and \texttt{PIM\_StBuf} for \emph{internal transfers} between the buffer and DRAM banks.} This division allows independent scheduling of requests, minimizing CPU memory access interference by using bus idle time windows for external transfers and exploiting bank idle time window for internal transfers. It enables the idleness-aware scheduling policy detailed in~\sectionautorefname~\ref{sec:scheduling:scheduler}.

The \texttt{PIM\_LdBuf} and \texttt{PIM\_StBuf} commands are designed to mirror the memory access behavior of \texttt{PIM\_Exec(Ld)} and \texttt{PIM\_Exec(St)}: they exclusively access the DRAM bank using internal bandwidth, transferring data from/to the bank to/from the buffer. and do not consume external bandwidth. Consequently, they inherit the same execution model: preemptable operation with a command length of $nPTL$. This design choice enables unified scheduling logic in all PIM-initiated bank activities.

\textbf{Additional PIM Buffer Requirement.}
To support our mechanism, we add an extra segment to the PIM buffer whose capacity is commensurate with the maximum amount of data transferred by a single \texttt{PIM\_LdBuf} or \texttt{PIM\_StBuf} command of length $nPTL$. 
\revision{Since internal bandwidth is frequently available during ample idle time windows, the scheduler can always complete these short internal transfers before the next external transfer command arrives.
Thus, a small per-bank buffer suffices to avoid stalling, as validated in our experiment (\sectionautorefname{}~\ref{sec:experiment:scheduling}).} 

\textbf{Timing Constraints.}
\tableautorefname{}~\ref{tab:timing_constrain} provides the timing constraints for the new commands. \texttt{PIM\_RdBuf} and \texttt{PIM\_WrBuf} commands only use the external memory channel bandwidth, requiring a $tBL$ delay after standard \texttt{Read}/\texttt{Write} commands on the same channel. These two commands, along with \texttt{PIM\_LdBuf}/\texttt{PIM\_StBuf}, need the PIM buffer to be ready before execution, adding a $tBL$ delay when issued to the same bank. Additionally, \texttt{PIM\_LdBuf} and \texttt{PIM\_StBuf} access the DRAM bank like \texttt{PIM\_Exec(Ld)} and \texttt{PIM\_Exec(St)}, following the same timing as DRAM commands like \texttt{ACT}, \texttt{PRE}, \texttt{Read}, and \texttt{Write}.

\textbf{Memory Ordering and Buffer Consistency.}
The bandwidth-decoupled transfer mechanism guarantees memory ordering and buffer consistency via strict execution sequencing.
Decoupled command pairs (\texttt{PIM\_RdBuf}/\texttt{PIM\_WrBuf} and \texttt{PIM\_LdBuf}/\texttt{PIM\_StBuf}), derived from single CPU-mediated data transfers, must execute in program order, enforced by the memory scheduling strategy in \sectionautorefname~\ref{sec:scheduling}.
\revision{
While other concurrent commands (PIM execution and CPU access) can be interleaved with the decoupled command pairs without ordering constraints since they never access the dedicated PIM buffer regions.}
These commands can arbitrarily interleave with the data transfer stream, even between decoupled pairs, without violating correctness.

\section{Idleness-aware Memory Scheduling}
\label{sec:scheduling}

This section describes our idleness-aware memory scheduling strategy designed to enhance PIM performance under the CPU-first scheduling principle. The strategy is implemented through two key mechanisms. First, IWE dynamically computes idle time windows for each DRAM bank and the memory bus by analyzing the CPU request queue. This enables more accurate scheduling decisions by the PIM scheduler and Command Arbiter. Second, utilizing the metadata provided by IWE, the PIM scheduler selects commands that optimize the utilization of these idle time windows in both the memory bus and the banks, while simultaneously maintaining low latency for CPU requests.

\subsection{Idle Window Estimater (IWE)}
\label{sec:scheduling:analyzer}

\subsubsection{Bank Idle Time Window Estimation}

The occurrence of idle time windows in DRAM banks can be attributed to two primary factors. First, the CPU is generally unable to maintain consecutive memory requests in rapid succession due to the sporadic nature of memory access at the application level, resulting in idle time windows between accesses to the same bank. These inter-request idle time windows have already been leveraged by previous CPU-first scheduling algorithms~\cite{Chopim9138972}. Such schedulers initiate PIM commands when the CPU request queue for a bank is empty and promptly suspend PIM operations upon the emergence of any new CPU request.

The second type of idle time window occurs when, despite the availability of multiple banks with open rows prepared for access, serialization of their access commands over a shared memory bus leads to periods of inactivity. Specifically, this inactivity occurs between the activation of a row (\texttt{ACT}) and its designated data access, during which the bank remains idle with the row open, resulting in the internal bandwidth waste highlighted in \figureautorefname{}~\ref{fig:scheduling-methods}(c).
This interval can be strategically optimized by delaying the \texttt{ACT} command until immediately before the data access. 
In the \archname{} architecture, the IWE module forecasts the earliest feasible service time for subsequent CPU access of each bank, based on the pending CPU requests.
Using this information, the IWE module instructs the Command Arbiter to postpone premature row activations. IWE module uses the resulting idle time window in the bank for PIM operations. As illustrated in \figureautorefname{}~\ref{fig:scheduling-methods}(e), the delayed row activation creates a window that can accommodate an additional PIM execution, improving the utilization of internal bandwidth.

IWE estimates the earliest time the next CPU request will be issued for each bank by simulating the FR-FCFS scheduling order of pending requests, thereby estimating each bank's future idle time window. 
Although the current prototype is designed for FR-FCFS scheduling, IWE module is intrinsically adaptable to different scheduling policies, via modifications to its estimation logic that align with alternative baseline scheduling policies (e.g., \cite{BLISS_6974655,atlas5416658}).
Because the request queue can dynamically change as new requests arrive or CPU memory accesses are completed, IWE must provide rapid predictions. 
Therefore, IWE leverages two key characteristics of FR-FCFS. 
First, row-hit requests are processed consecutively, since they bypass row activation and only require the memory bus (i.e., the external bandwidth), preempting requests from other banks. Secondly, requests within the same rank are grouped to prevent rank switch penalties ($tRTRS$), deferring inter-rank switching until there are no more ready requests in the current rank. These insights allow IWE to closely approximate the actual scheduling order with small overhead, forming the foundation of our algorithm that estimate the earliest access cycle of each bank (Algorithm~\ref{alg:prediction}).

\begin{algorithm}[h]
\caption{Earliest Access Cycle Estimation in IWE}
\label{alg:prediction}
\begin{algorithmic}[1]
    \Require $REQ[]$ (Earliest-arriving request of each bank)
    \State $ready\_cycles$ $\gets$ [$get\_ready\_cycle(r)$ for $r$ in $REQ$]
    \State $t \gets cur\_tick()$, $cr \gets cur\_rank()$, $service\_time \gets \{\}$
    \While {$REQ.size()$}
        \State $AnyReady \gets Any([r.rank == cr$ \&\& $r.ready \leq t$ for $r$ in $REQ])$
        \If{$AnyReady$} 
            \State $r \gets earliest\_ready([r$ for $r$ in $REQ$ if $r.rank==cr])$, $t \gets t + tBL$
        \Else
            \State $r \gets$ $earliest\_ready(REQ)$
            \State $t \gets \text{max}(r.ready,t)$, $cr \gets r.rank$
        \EndIf
        \State $service\_time[r] \gets t$, $ REQ.remove(r)$
    \EndWhile
    \For {$b$ in $range(bank\_num)$}
        \State $window\_bank[b] \gets service\_time[r]~\vert~r.bank==b$
    \EndFor
    \State $window\_bus \gets$ min $(service\_time.values)$ 
    \State \Return $window\_bank$, $window\_bus$ 
\end{algorithmic}
\end{algorithm}

\revision{
In this algorithm, IWE estimate the earliest service time of each bank's earliest-arriving request to obtain the idle windows.
In line 1, IWE calculate the ready cycles of each bank's earliest-arriving request ($REQ[]$) according to the bank state.
Each bank can be in one of the three states: (1) \emph{Row-Closed}, which requires an \texttt{ACT} command, delaying the access by at least $tRCD$ from the current cycle; (2) \emph{Opened-to-the-target-row}, which permits the access $tRCD$ after the row's opening cycle; (3) \emph{Opened-to-a-different-row}, which requires both \texttt{PRE} and \texttt{ACT} commands, deferring the access past $tRP + tRCD$ from the current cycle.
In line 2, the algorithm initializes $t$ to current tick and $cr$ to the rank of the previously issued command.
Then, in the loop, it select the earliest ready request in rank $cr$ at tick $t$ for issuing, and advances $t$ by $tBL$ (line 6). If none is found, a rank switch occurs: the earliest request from the whole list is chosen, updating $cr$ to its rank (lines 8-9). After processing all requests, IWE determines each bank's idle time window according to the earliest service cycle of each bank (line 12).
}

\textbf{Idleness-Aware Command Arbitration.}
The Command Arbiter utilizes idle time window estimates from IWE to guide two primary choices for command scheduling decisions. 
First, when the CPU memory scheduler issues a \texttt{ACT} command, the Command Arbiter evaluates if the estimated idle time window for a bank is long enough for a PIM operation to take place. This window needs to encompass the time required for row switching ($tRP+tRCD$) and must also permit the PIM unit to process at least one column. If these conditions are met, the \texttt{ACT} command is deferred to allow for PIM execution. 
Second, if the CPU request queue for a bank becomes non-empty while a PIM unit is executing, the Command Arbiter refrains from issuing a \texttt{PIM\_Pause} immediately. Rather, it delays the pause until the last possible cycle that would ensure the next CPU memory access is not delayed, maintaining the CPU-first principle.

\subsubsection{Memory Bus Idle Window Estimation}

\revision{
Alongside estimating idle time windows for each bank, IWE estimates idle time windows on the memory bus. This allows the PIM scheduler to issue external transfer PIM commands (\texttt{PIM\_RdBuf} and \texttt{PIM\_WrBuf}), without hindering CPU accesses. 
The estimation utilizes the earliest service times of all outstanding CPU requests (Line 13 of Algorithm~\ref{alg:prediction}). 
The memory bus will remain idle from the present cycle until cycle $window\_bus$. As a result, any external transfer PIM command that can be completed before this cycle can be issued without delaying CPU access.}

\subsection{PIM Scheduler}
\label{sec:scheduling:scheduler}

\begin{figure}[h]
    \centering
    \includegraphics[width=0.95\linewidth]{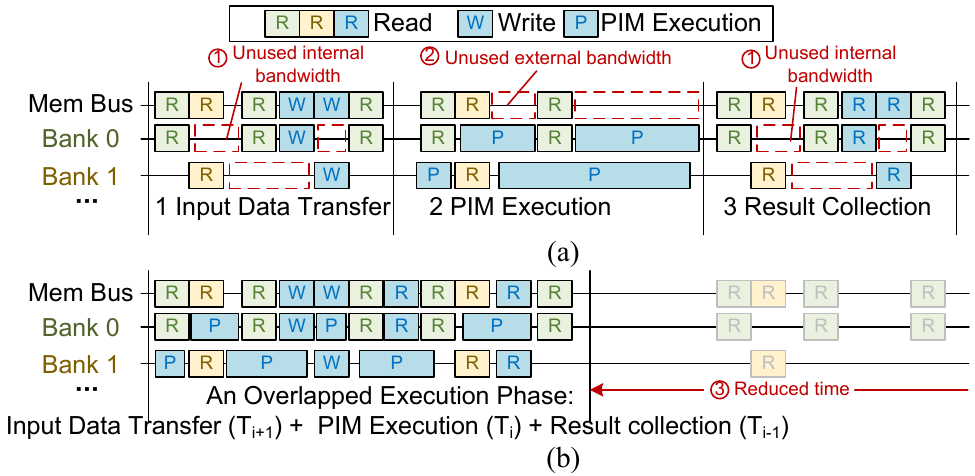}
    \caption{(a) Conventional software-controlled three-stage sequential scheduling. (b) Overlapped scheduling. The row open operations are omitted for visual simplicity.}
    \label{fig:Overlapped}
\end{figure}

The PIM scheduler dynamically chooses the best PIM command from PRWQ and PEQ to utilize idle time windows. In conventional PIM execution, depicted in \figureautorefname{}~\ref{fig:Overlapped}(a), the process includes three sequential stages: input data transfer, PIM execution, and result collection, all managed by the CPU software sequentially. This sequential processing prevents simultaneous usage of both internal bandwidth (\circnum{1}) and external bandwidth (\circnum{2}).

To avoid this sequential execution and improve both internal and external bandwidth usage, we propose an \emph{overlapped scheduling strategy} within the PIM scheduler. Unlike the traditional method where each stage must finish entirely before the next can begin, our strategy divides the PIM workload into loosely-coupled tiles (e.g., submatrices in matrix multiplication). This allows CPU-mediated transfers of one tile to coincide with the PIM execution of another, since there are no data dependencies between the tiles (the CPU handles the reduction). As shown in \figureautorefname{}~\ref{fig:Overlapped}(b), the PIM scheduler issues compatible commands from different tiles \emph{simultaneously} (e.g., collecting results for tile $T_i$ while executing tile $T_{i+1}$), creating an \emph{overlapped execution phase}. This approach maximizes the use of both internal and external bandwidth, thereby enhancing the end-to-end performance of PIM workloads (\circnum{3}). To guarantee that data dependencies are satisfied, a \texttt{PIM\_Barrier} command is introduced between the overlapped execution phases, ensuring that all operations of one phase are completed before the next phase starts.

During each overlapped execution phase, the PIM scheduler chooses a PIM command from the PRWQ or the PEQ, based on a priority policy. The top priority is to issue a \texttt{PIM\_RdBuf} or \texttt{PIM\_WrBuf} command from PRWQ, as these commands solely occupy the external memory bus, which is often the bottleneck. Hence, it is crucial to efficiently utilize such external bus idle times. If no bus command is ready, the scheduler gives precedence to \texttt{PIM\_LdBuf} and \texttt{PIM\_StBuf} to ensure that the data is promptly loaded from or stored into the PIM buffer, preventing the blockage of subsequent bus commands.
Note that our scheduler preserves arrival order for PRWQ commands targeting the same bank, correctly handling any ordering requirements between different requests. 
Specifically, when the reserved buffer segment is full during writing, the next issued command is necessarily \texttt{PIM\_StBuf} that clears the buffer. 
Only when no PRWQ commands can be sent will the scheduler opt for a PIM execution command from the PEQ that matches the current idle period of the bank.

\section{Effectiveness on the Software Stack}

The \archname{} framework maintains programmer transparency through compiler-assisted command translation while maintaining backward compatibility with existing PIM programs that target command-driven PIM architecture \cite{AiMSW9895629,HBM-PIM-SW9499894}. 
Conventional PIM kernels (e.g., MAC and softmax) are directly translated into preemptable execution commands (PIM\_Exec(L/S)) without modifying user-level code or computational semantics.
CPU-mediated data transfers are automatically decomposed into adjacent command pairs: write sequences use \texttt{PIM\_WrBuf} \& \texttt{PIM\_StBuf} while read sequences employ \texttt{PIM\_LdBuf} \& \texttt{PIM\_RdBuf} (\sectionautorefname{}~\ref{sec:interface:decoupled}). 
To enable overlapped scheduling, the compiler automatically inserts \texttt{PIM\_Barrier} commands at tile boundaries, ensuring the correctness of concurrent execution of data transfers and computations across independent tiles.

As our solution requires no changes to program code or OS, no additional data-transfer overhead is incurred.
This preserves the inherent advantage of LLM inference workloads, where weight matrices are pre-organized in banks during model loading and remain static throughout inference, following established practices in PIM-accelerated systems \cite{PAPI10.1145/3676641.3716009,attacc10.1145/3620665.3640422}. 
The primary change \archname{} introduces occurs at the memory controller driver level, which requires support for the new command semantics and scheduling interfaces described in \sectionautorefname~\ref{sec:interface}.

\section{Evaluation}
\label{sec:evaluation}

\subsection{Experimental Setup}
\label{sec:evaluation:setup}

\textbf{Simulation.} The performance of \archname{} and baselines is evaluated with the Ramulator2 simulator \cite{ramulator2_10321647,ramulator2_github} (based on Ramulator \cite{ramulator1_7063219}).
The PIM commands and related timing constraints to the modeled LPDDR5 module.
We extend the memory controller model with the additional modules described in \sectionautorefname{}~\ref{sec:architecture}.
We use DRAMPower \cite{drampower10.1145/3721848.3721850} to evaluate the power consumption of DRAM.
\tablename{}~\ref{tab:configure} summarizes the modeled system configuration.
The host CPU configuration is based on a mobile phone with a Qualcomm Snapdragon 888 \cite{qualcomm_snapdragon888} and 32GB of DRAM memory.
The PIM unit to DRAM bank bandwidth and energy consumption are modeled on a taped-out PIM chip \cite{samsunglpddr5pim10254711}.
We use the O3CPU front-end model of Ramulator2 \cite{ramulator2_10321647,ramulator2_github} for CPU simulation, with the CPU workload input to the front-end as a memory trace.
For CPU workloads from SPEC CPU2017 and PolyBench-ACC, we use the zsim simulator \cite{zsim10.1145/2485922.2485963} to generate the memory traces.
A memory trace contains the memory accesses and the corresponding latency between the accesses.
For mobile phone applications, we use the Xiaomi Mi 11 Pro \cite{xiaomi_mi11pro_specs} smartphone to run the workload and collect memory traces with instrumentation tools, Frida \cite{frida_stalker}.
The Xiaomi Mi 11 Pro is equipped with a Qualcomm Snapdragon 888 SoC \cite{qualcomm_snapdragon888} (1 x 2.84 GHz Cortex-X1 + 3 x 2.42 GHz Cortex-A78 + 4 x 1.8 GHz Cortex-A55), 12GB LPDDR5 RAM \cite{JEDECLPDDR52023}, and 256GB storage, running Android 15 \cite{android15_developer}.
The DRAM configuration follows LPDDR5-6400 standards \cite{JEDECLPDDR52023}.
\revision{The CPU memory traces are repeatedly replayed to generate background DRAM traffic.}
To evaluate \archname{}'s area overhead, we synthesize the hardware modules in the memory controller using the Synopsys Design Compiler with TSMC 90nm technology library at the frequency of 2.4GHz.
Since modern smartphone SoCs are typically fabricated in processes below 5nm, we provide a conservative area estimate for a 5nm implementation by scaling logic and memory components separately, based on publicly available technology scaling data \cite{5nm8993577}.

\begin{figure*}[hb]
    \centering
    \includegraphics[width=0.95\linewidth]{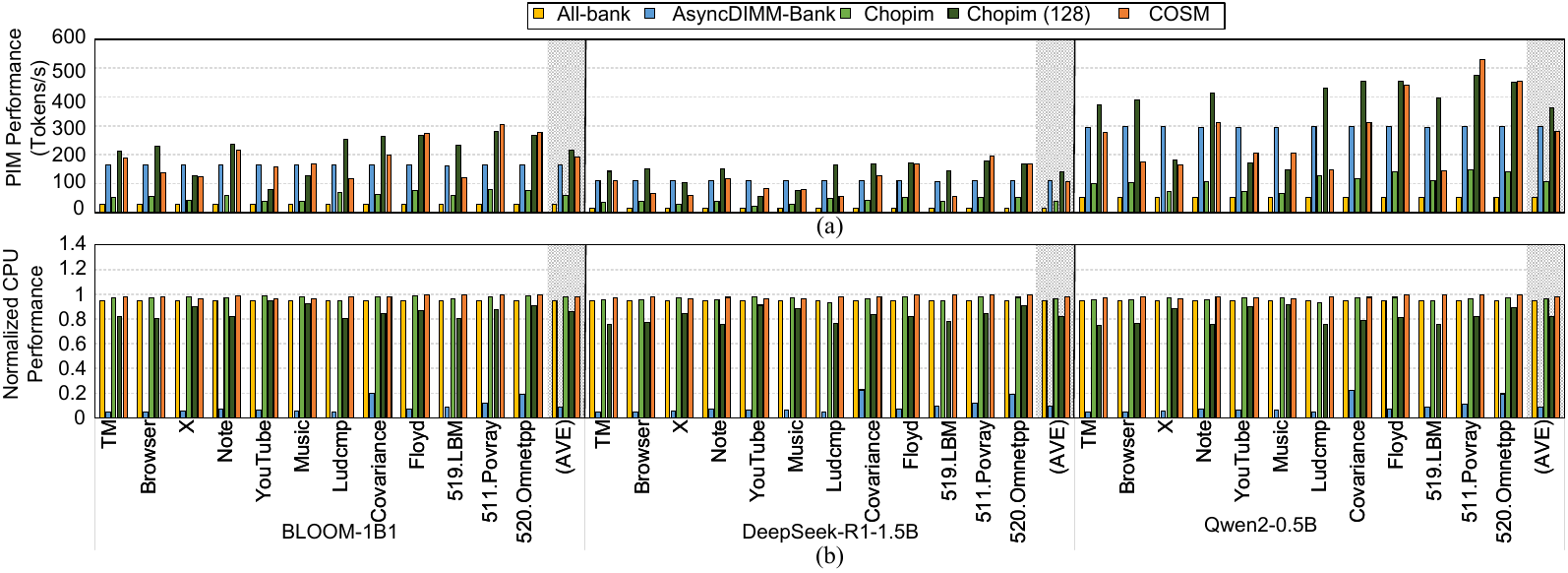}
    \caption{Overall PIM \& CPU performance of \archname{} and baselines for concurrent CPU and PIM execution.}
    \label{fig:exp1_overall_performance}
\end{figure*}

\begin{table}[h]
    \centering
    \caption{System Configuration}
    \label{tab:configure}
    \resizebox{0.9\columnwidth}{!}{
    \begin{tabular}{cl}
        \toprule
        \multicolumn{2}{c}{Host CPU }                                       \\\midrule
        Processor           & 8 \texttimes{} CPU cores @2.1GHz on average         \\
                            & Out of Order, 160 entry RoB, 4 IPC                  \\
        L3                  & 4MB, Assoc: 8, 64B Cache Line Size             \\
        CPU Scheduler    & FR-FCFS \cite{FR-FCFS10.1145/339647.339668,zuravleff1997controller}                                       \\
        PIM Queue           & PEQ size: 2, PRWQ size: 2 (per bank)  \\
        \bottomrule
        \toprule
        \multicolumn{2}{c}{ DRAM }                                     \\\midrule
        DRAM                & LPDDR5-6400, 8GB/Rank                         \\
        Organization   & Bank / Bank Group / Row / Column                        \\
         & 4  / 4 / 16384 / 2048B                        \\
        Timing Param.   & tBL=2.5 (16), tRCD=tRP=4.7, tCL=6.3,                  \\
        (ns)        &  tRAS=10.7, tRRD=1.3,  tRFC=87.5, tWR=8.8,                        \\
                    &  tWTR=3.1, tRTP=1.3, tCS=0.6, tREFI=967.5                  \\
        \bottomrule
        \toprule
        \multicolumn{2}{c}{ PIM Units }                                     \\\midrule
        PIM Core            & 1GHz, 6.4 GB/s bandwidth 6.4TFLOPS \cite{samsunglpddr5pim10254711}          \\
                            & 1kB Buffer for CPU-mediated transfer \\
                            & 16-bit PIM-bank wire width        \\
        Num                 & 16 per Rank, at Bank level                    \\
        \bottomrule
        \toprule
        \multicolumn{2}{c}{ System Configuration }                          \\ \midrule
        CPU System          &  2 Channels \texttimes{}2 Ranks with PIM units    \\
        \bottomrule
            
    \end{tabular}
    }

\end{table}

\textbf{CPU Benchmarks.}
We evaluate \archname{} using six mobile applications, three PolyBench benchmarks \cite{Polybench-acc6339595}, 
and three SPEC CPU2017 \cite{spec2017} specifically selected for their diverse memory access patterns.
We use the following application workloads:
\begin{itemize}
    \item \emph{Tencent Meeting (TM)}: Video conference app; trace collected during whiteboard sharing.
    \item \emph{Browser}: Default mobile web browser; trace collected during page loading.
    \item \emph{X}: Social network; trace collected when clicking an article.
    \item \emph{Note}: Note-taking app; trace collected during typing.
    \item \emph{YouTube}: Video streaming platform; trace collected during video watching.
    \item \emph{Music}: System music streaming platform; trace collected during song playback.
\end{itemize}
PolyBench benchmarks show dense, contiguous memory access, leading to high memory bandwidth utilization and high row hit rates, while SPEC CPU2017 targets broader testing.
\tableautorefname{}~\ref{tab:rhr} present the row hit rate of the evaluated benchmarks.

\begin{table}[h]
    \centering
    \caption{Row Hit Rate of Benchmarks.}
    \label{tab:rhr}
    \resizebox{0.85\linewidth}{!}{
    \setlength{\tabcolsep}{4pt} 
    \begin{tabular}{cc|cc|cc}
    \toprule
    Bench. & Rate & Bench. & Rate & Bench. & Rate \\ \midrule
    TM & 0.039 & YouTube & 0.187 & Floyd & 0.908 \\
    Browser & 0.434 & Music & 0.010 & 519.LBM & 0.843 \\
    X & 0.056 & Ludcmp & 0.870 & 511.Povray & 0.596 \\
    Note & 0.011 & Covariance & 0.001 & 520.Omnetpp & 0.958 \\\bottomrule
    \end{tabular}
    }
\end{table} 

\textbf{PIM Benchmarks.}
We tested inference on mobile devices using three open-source LLMs, each with around one billion parameters: BLOOM-1B1 \cite{bigscience2022bloom}, DeepSeek-R1-1.5B \cite{deepseekai2025deepseekr1incentivizingreasoningcapability}, and Qwen2-0.5B \cite{qwen2}. BLOOM-1B1 is a multilingual model, DeepSeek-R1-1.5B specializes in programming, and Qwen2-0.5B is a compact bilingual model. Despite differences, all are suitable for mobile deployment. Benchmarks use 16-bit quantized inputs and 8-bit quantized weights.

\textbf{Baselines.}
We compare \archname{} that use three baselines with different PIM control interfaces and scheduling strategies: \emph{All-Bank Command} \cite{HBM-PIM-HW9365862}, \emph{Chopim} \cite{Chopim9138972}, and \emph{AsyncDIMM-Bank}, which is the bank-level PIM version of the original \emph{AsyncDIMM} \cite{AsyncDIMM10946818}. 
In \emph{All-Bank Command}, PIM operations are issued using all-bank commands that simultaneously invoke all PIM units.
Since CPU memory access is blocked after issuing the all-bank PIM commands, we introduce a time-sliced round-robin strategy: assume that 95\% of the time is spent on CPU memory access, while the remaining 5\% is allocated to PIM computation: we consider CPU performance degradation to be small enough as long as it is smaller than a 5\% threshold. 
Crucially, to ensure a fair and conservative comparison, we assume idealized zero-overhead switching between PIM and CPU phases in this baseline. 
This isolates the impact of the memory interface architecture on scheduling potential.
In contrast, \emph{Chopim} and \emph{AsyncDIMM-Bank} adopt single-bank commands with the PIM execution command length equal to $tBL$. \emph{Chopim} adopts a CPU-first scheduling strategy, which prioritizes CPU memory accesses over PIM computation by blocking the PIM command queue whenever the CPU memory queue of a bank is not empty. 
\emph{AsyncDIMM-Bank} adopts a relatively fair strategy to maximize the row hit rate by switching between the PIM command queue and CPU memory queue upon detecting a \texttt{PRE} command for either CPU or PIM unit. 
Moreover, it uses a relay memory controller within the rank to reduce command bus pressure. None of the baselines mentions their scheduling strategy of CPU-mediated data transfer. We assume that these memory accesses are scheduled together using the FR-FCFS policy. \archname{} adopts both the low-interference PIM control interface and the idleness-aware scheduling strategy. We set the PIM execution command length to 128 cycles, a value that balances the command bus contention (if too short) against excessive \texttt{PIM\_Pause} commands and their bus bandwidth consumption (if too long), as observed in our experiments.

\begin{figure*}[h]
    \centering
    \includegraphics[width=0.95\linewidth]{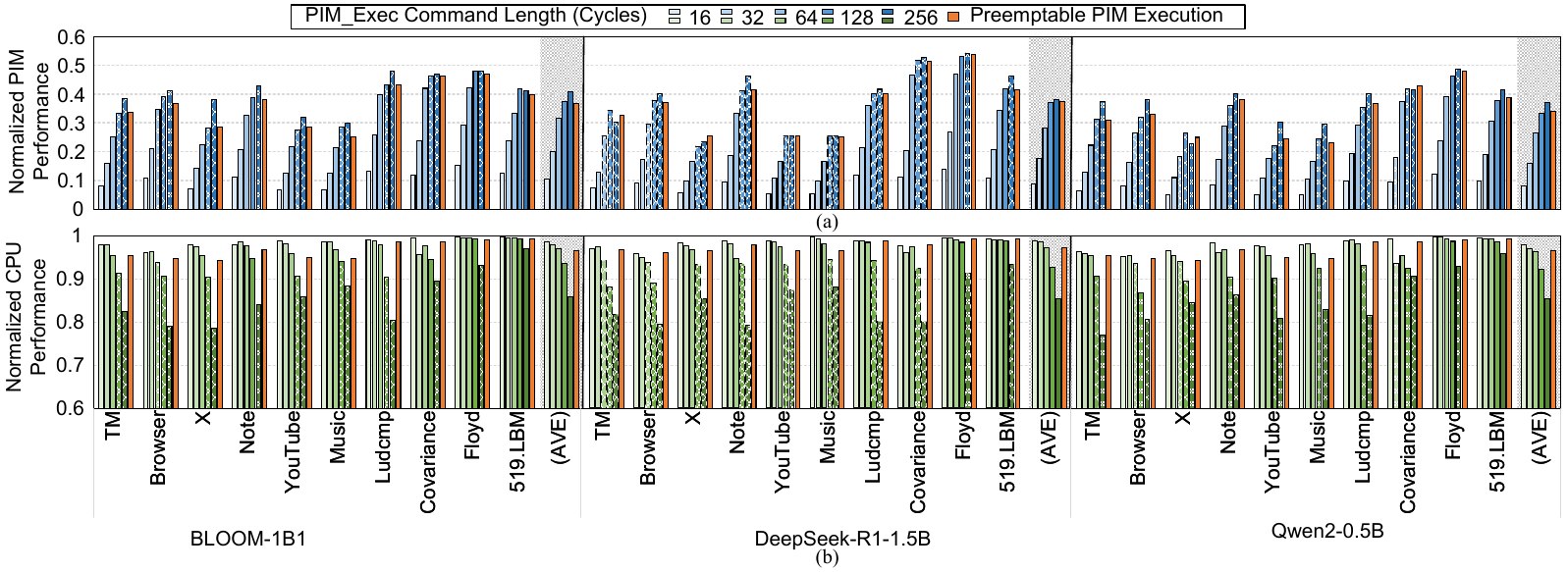}
    \caption{Normalized CPU and PIM workload performance under fixed-length and preemptable PIM execution command. Cases where CPU performance degrades by more than 5\% are marked with dashed lines.}
    \label{fig:exp2_pim_task_len}
\end{figure*}

\subsection{Overall Performance}
\label{sec:evaluation:performance}

\revision{
\figureautorefname{}~\ref{fig:exp1_overall_performance} shows the performance of \archname{}'s CPU and PIM benchmarks during simultaneous execution compared to the baselines. 
CPU performance is normalized to the performance obtained under standalone execution, which indicates the ideal case.}
The key-value cache is configured to a length of 2k. 
We also include a baseline called \emph{Chopim(128)} by increasing the PIM execution command length of \emph{Chopim} to 128 cycles.
For the \emph{All-bank} baseline, the PIM units are restricted to using only 5\% of the time windows, missing the opportunity to take advantage of the idle periods between CPU memory accesses. 
The CPU-first approach of \emph{Chopim} results in a small 3.0\% slowdown for the CPU but only provides a relatively small 1.9\texttimes{} increase in PIM throughput over \emph{All-Bank}, due to command-bandwidth contention. 
\emph{AsyncDIMM-Bank} provides a 4.2\texttimes{} PIM throughput compared to \emph{All-bank}, yet it significantly affects CPU performance, reducing it by an average of 89.9\%. 
Although increasing the PIM execution command length to 128 cycles can increase the PIM throughput of \emph{Chopim} by 3.44\texttimes, the CPU performance reduction increased to 13.5\%, indicating the trade-off between PIM performance and CPU performance for fixed-length commands.
In contrast, the scheduling strategy of \archname{} maintains CPU performance with only a 2.0\% degradation. 
Additionally, the low-interference PIM control interface and idleness-aware scheduling enable \archname{} to more effectively use the remaining internal bandwidth for PIM tasks, leading to a 6.0\texttimes{} improvement over the \emph{All-Bank} baseline and 2.8\texttimes{} over \emph{Chopim}.

\subsection{Effect of Preemptable PIM Execution}

\figureautorefname{}~\ref{fig:exp2_pim_task_len} \revision{illustrates the performance of CPU and PIM benchmarks during concurrent execution, comparing traditional fixed-length commands across different lengths (from 16 to 256), against our proposed preemptable PIM execution command.}
\revision{Both PIM and CPU performance is also normalized to performance of standalone execution.}
The maximum command length of 256 remains below the latency threshold for processing a complete row in our setup. 
The PIM workload corresponds to a KQV generation layer from the three models that is dominated by PIM execution.
For fixed-length commands, there is a clear trade-off between CPU and PIM performance: longer commands enhance PIM workload throughput but negatively impact CPU performance. Notably, when the PIM execution command length surpasses 64 cycles, PIM performance reaches saturation in most cases due to decreased command bus pressure. This phenomenon is aligned to \figureautorefname{}~\ref{fig:ob-cpu-lat-and-pim-len}(b).
However, CPU performance significantly declines by over 5\%, particularly in mobile applications with random access patterns. 
In contrast, our preemptable interface can achieve maximum PIM performance without with only 3.2\% degradation in CPU performance. 
Compared to a fixed command length of 32 cycles, which maintains CPU interference below 5\% for all workloads, our preemptable PIM execution design results in a 2.02\texttimes{} improvement in PIM performance.

\begin{figure*}[ht]
    \centering
    \includegraphics[width=0.95\linewidth]{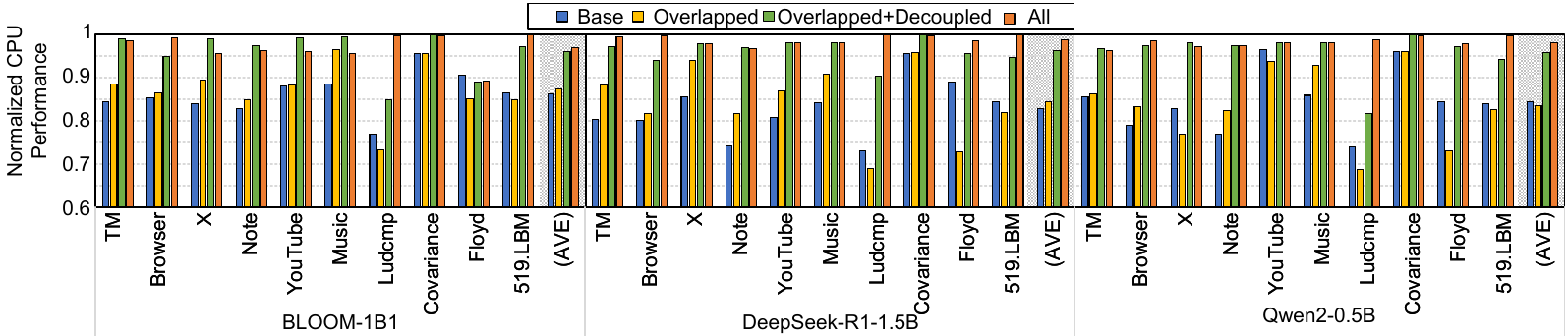}
    \caption{Impact of CPU-mediated data transfers on CPU performance under different scheduling strategies. We test on the attention layers of the benchmarks.}
    \label{fig:exp3_data_transfer}
\end{figure*}

\subsection{\revision{Sensitivity Analysis on CPU Performance during CPU-mediated Data Transfer}}

To assess how our techniques affect CPU-mediated transfers, we evaluate CPU performance normalized to the standalone execution, under four different configurations using attention layers from the three models that are characterized by substantial CPU-mediated transfer, as depicted in \figureautorefname{}~\ref{fig:exp3_data_transfer}.
To assess how our techniques affect CPU-mediated transfers, we evaluate the normalized CPU performance under four configurations using attention layers from three models characterized by substantial CPU-mediated transfers, as depicted in \figureautorefname{}~\ref{fig:exp3_data_transfer}.
In the \emph{Base} configuration, CPU-mediated transfers occur through standard read/write requests, following the traditional three-stage PIM execution. 
\emph{Overlapped} execution facilitates the simultaneous scheduling of CPU-mediated transfers and PIM execution commands by utilizing our scheduler design described in \sectionautorefname{}~\ref{sec:scheduling:scheduler}. This approach increases CPU performance by an average of 3.7\% across mobile workloads compared to \emph{base}. 
However, for workloads with high row-hit rates (such as algebraic kernels in PolyBench-ACC), this aggressive overlap can inadvertently cause additional row switches, leading to greater CPU interference. 
\revision{The \emph{Overlapped+Decoupled} configuration further employs bandwidth-decoupled CPU-mediated PIM data transfer commands (\sectionautorefname{}~\ref{sec:interface:decoupled}), which reduces contention with CPU accesses and increase CPU performance by 11.5\% compared to \emph{Base}.} 
Finally, the \emph{All} configuration utilizes IWE to proactively detect memory bus idle time windows that are sufficiently large to accommodate entire burst transfers, thereby maintaining CPU slowdown consistently below 5\% across all evaluated workloads.

\subsection{Effect of Idleness-aware Scheduling on PIM Execution}
\label{sec:experiment:scheduling}

\begin{figure}[h]
    \centering
    \includegraphics[width=0.85\linewidth]{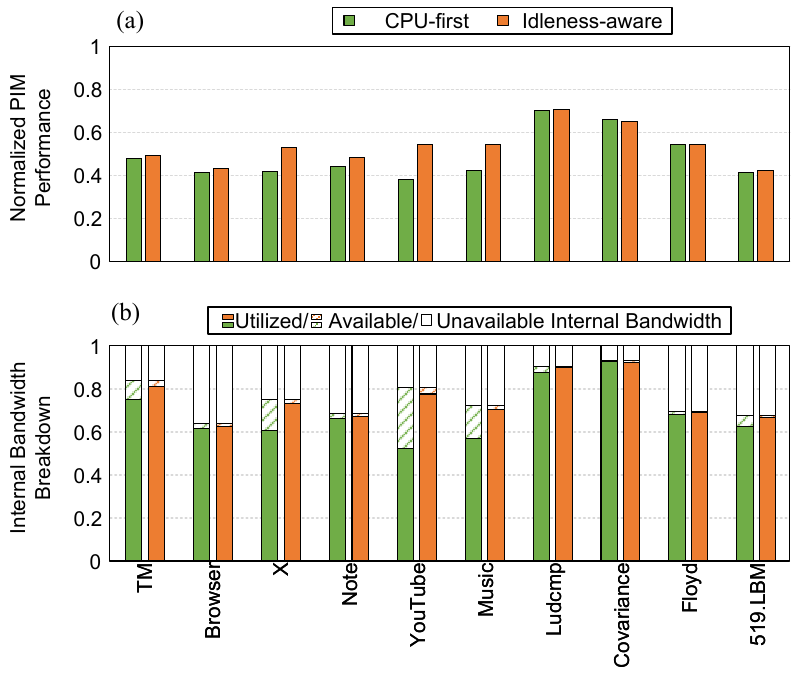}
    \caption{(a) PIM performance and (b) Internal bandwidth usage during concurrent execution with CPU workload under CPU-first scheduling and idleness-aware scheduling.}
    \label{fig:exp4_idle_aware}
\end{figure}

In \figureautorefname{}~\ref{fig:exp4_idle_aware}(a), we illustrate the performance of PIM workloads when managed by CPU-first scheduling compared to idleness-aware scheduling (\sectionautorefname{}~\ref{sec:scheduling}) when running alongside with CPU workloads, with all results normalized to their standalone execution. The PIM workload is derived solely from the execution segment of an attention layer, excluding CPU-mediated data transfer, from the DeepSeek model. 
Under CPU-first scheduling, the issuance of PIM execution commands halts if a bank queue is not empty. 
In the idleness-aware configuration, idle time windows are predicted using IWE to enhance performance. 
This configuration provides an average performance boost of 1.21\texttimes{} for PIM workloads across mobile applications. \figureautorefname{}~\ref{fig:exp4_idle_aware}(b) displays the bandwidth usage under both scheduling strategies. The term ``available bandwidth'' refers to idle time windows that are long enough for a PIM execution command but are not used due to scheduling limitations, whereas ``unavailable bandwidth'' pertains to short, fragmented idle periods unsuitable for PIM execution. Compared to CPU-first scheduling, Idleness-Aware scheduling exploits an additional 37.0\% of the available bandwidth, leaving less than 1\% unused.

\subsection{Energy Consumption}

\begin{figure*}[hb]
    \centering
    \includegraphics[width=0.95\linewidth]{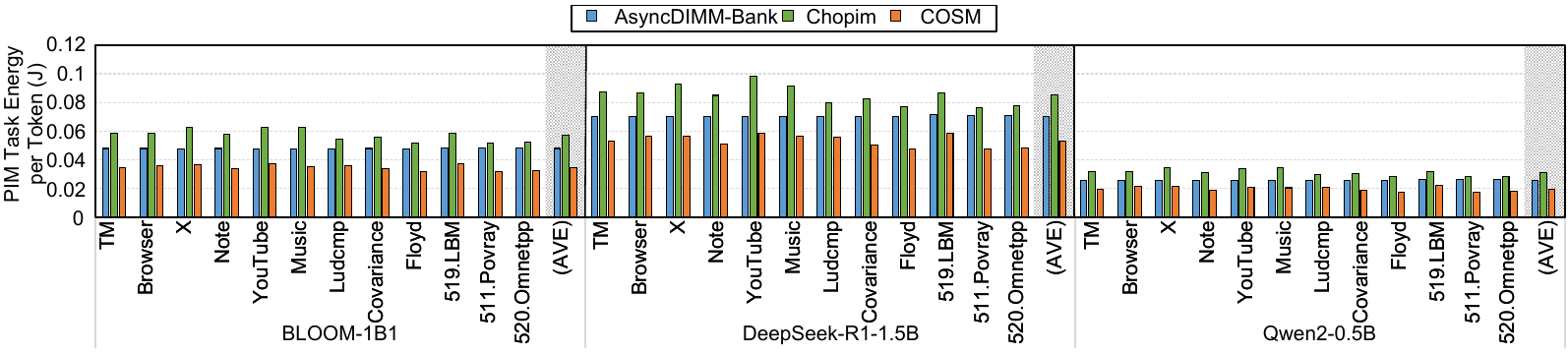}
    \caption{PIM workload energy consumption per token (including PIM unit computation, PIM bank access, and CPU-mediated data transfer) of \archname{} and baselines during concurrent CPU and PIM execution.}
    \label{fig:exp5_energy}
\end{figure*}

\figureautorefname{}~\ref{fig:exp5_energy} depicts energy consumption per token of \archname{} and baselines.
We only calculate the energy of PIM computation, and exclude the energy of CPU access during concurrent execution.
Although \archname{}'s fine-grained CPU-PIM task interleaving increases row switching overheads, it achieves a reduction in total energy by eliminating long active-idle periods. 
\archname{} reduces energy by 1.34\texttimes{}/1.61\texttimes{} compared to the AsyncDIMM-Bank/Chopim baseline.

\subsection{Sensitivity Analysis}

\begin{figure}[!htbp]
    \centering
    \includegraphics[width=0.95\linewidth]{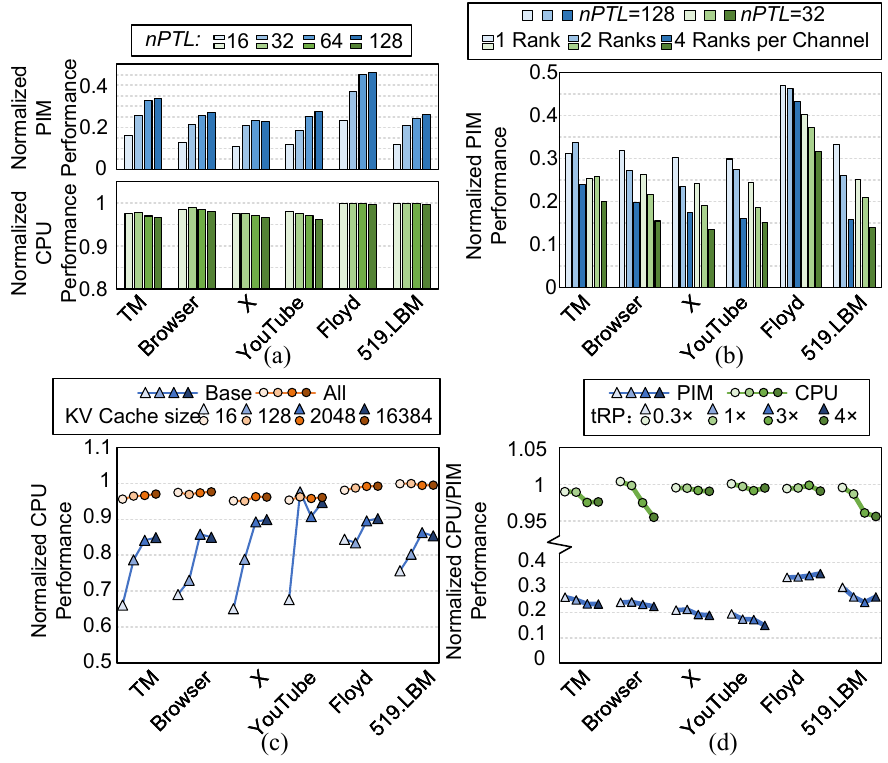}
    \caption{Sensitivity analysis on (a) $nPTL$ (b) rank count per channel (c) KV cache size (d) scaled $tRP$ ($n\times$, relative to \tableautorefname{}~\ref{tab:configure} configuration).}
    \label{fig:exp6_sa}
\end{figure}

\revision{\textbf{PIM Execution Command Length (\textit{nPTL}).} \figureautorefname{}~\ref{fig:exp6_sa}(a) analyzes the sensitivity of \archname{} to the $nPTL$ timing constrain. Generally, smaller command length ($nPTL<32$) saturates the command bus, exacerbating contention between CPU and PIM tasks. 
However, moderately reducing $nPTL$ can yield benefits by decreasing \texttt{PIM\_Pause} injections that potentially stall CPU accesses. 
For instance, in workload \textit{X}, reducing $nPTL$ from 128 to 64 improves both CPU and PIM performance.
}

\revision{
\textbf{Rank Count per Channel.}
\figureautorefname{}~\ref{fig:exp6_sa}(b) analyzes the PIM performance of \archname{} to the rank count in each channel.
Higher rank counts require the memory controller to issue more commands to trigger the execution of all PIM units, making the command bus easily saturated.
Therefore, scaling $nPTL$ from 32 to 128 yields a higher speedup for 4 ranks (1.27\texttimes{}) than for 1 rank (1.22\texttimes{}).}

\revision{
\textbf{KV Cache Size.}
\figureautorefname{}~\ref{fig:exp6_sa}(c) depicts the CPU performance degradation of \archname{} to the KV cache size of the DeepSeek attention layer, adopting the same series labels as \figureautorefname{}~\ref{fig:exp3_data_transfer}.
As the KV cache grows, CPU-mediated data transfer increasingly dominates the total computation time.
This mitigates the performance drop in the \emph{Base} case, which is highly sensitive to data transfer overhead. 
In contrast, \archname{} effectively manages both data transfer and PIM execution, maintaining CPU performance degradation below 5\%.
}

\revision{
\textbf{\textit{tRP}.} \figureautorefname{}~\ref{fig:exp6_sa}(d) shows performance under varying $tRP$ (row precharge time) using a DeepSeek attention layer, excluding CPU-mediated data transfer. 
Under the current timing configuration (listed in \tableautorefname{}~\ref{tab:configure}), the CPU-PIM interference remains minimal. 
Since CPU and PIM requests typically map to different rows, a larger $tRP$ heightens the row-switch penalty between CPU access and PIM execution command. 
This amplifies interference and degrade both CPU and PIM performance.
}

\subsection{Area Overhead.}
\label{sec:evaluation:area}
Implemented in a Snapdragon 8-class LPDDR5 memory controller (estimated at 0.93 mm², measured from publicly available die photographs), the \archname{} hardware modules occupy 0.069 mm², resulting in a modest 7.4\% area overhead.
This includes 0.014 $mm^2$ for the PIM scheduler, 0.0085 $mm^2$ for the IWE, and 0.0054 $mm^2$ for the Command Arbiter.


\section{Discussion}
\label{sec:discussion}

\textbf{Applicability to Broader Architectures.}
Although evaluated on mobile platforms, \archname{}'s idleness-aware scheduling is applicable to any shared-memory CPU-PIM system. 
As PIM extends to desktops, servers, data centers, and supercomputers for diverse workloads \cite{pushtap10.1145/3676642.3736120, attacc10.1145/3620665.3640422,FALA10.1145/3725843.3756127,primer_Mutlu2023,enabling_8405955}, they inherently encounter the same bank and bus contention challenges appear when CPU and PIM units contend with each other.
\archname{} addresses this through a universal memory abstraction: PIM workloads are characterized as bandwidth-sensitive, whereas CPU workloads are latency-sensitive

\textbf{Extension to Heterogeneous Accelerators.}
Although \archname{} targets CPU-PIM coordination, supporting heterogeneous agents (e.g., GPU/NPU) is vital. 
With memory-bound tasks offloaded to PIM, these accelerators handle compute-intensive tasks and remain latency-sensitive. 
\archname{}'s principle applies directly: located in the memory controller, it treats accelerator accesses as standard host requests, effectively isolating them from PIM-induced delays. 
Given the unique access patterns and scheduling policies of accelerators, extending \archname{} to explicitly capture these characteristics is a key research direction for future work.

\textbf{Compatibility with SIMD PIM Architectures}
While \archname{} currently targets single-bank PIM, its core principles readily extend SIMD PIMs (e.g., bank-group- or subarray-level) \cite{ambit_8686556,SIMDRAM_10.1145/3445814.3446749,mimdram_10476478,proteus_10.1145/3721145.3730420,drisa_8686604}.
However, the inherent parallelism of SIMD execution poses challenges to idle time window prediction, as it requires the simultaneous availability of multiple banks.
Consequently, future adaptations must balance PIM parallelism with CPU latency guarantees. 
Our scheduler enables this extensibility, but a dedicated exploration is left for future work.

\textbf{Thermal Considerations.}
While \archname{} maximizes PIM throughput by exploiting CPU idle time windows, it can support thermal management.
First, Command Arbiter can throttle PIM if the temperature exceeds thresholds.
Second, IWE can reduce idle time window utilization, explicitly trading PIM throughput for thermal safety.
Future work could explore adaptive strategies driven by real-time thermal feedback.
\section{Related Work}

\textbf{PIM Architecture.}
\revision{
PIM \cite{primer_Mutlu2023} integrates computational units inside memory devices to improve the performance of memory-intensive workloads.
This design philosophy applies universally across the entire storage hierarchy, including SRAM cache \cite{compute_caches_7920849,affinity10411388,infinity-stream10.1145/3582016.3582032,near-stream9773240,dalorex10071089,neuralcache_8416842,lut_9251975}, DDR \cite{CoNDA8980366,upmem8875680,axdimm10.1145/3533737.3535093,google_nn_2021_9563028,google_workload_10.1145/3173162.3173177}, GDDR\cite{AiMHW9731711}, HBM \cite{HBM-PIM-HW9365862}, 3D stacked memory \cite{H2-LLM10.1145/3695053.3731008,3D-PATH10.1145/3725843.3756087,scalable_pim_10.1145/2872887.2750386}, and SSD \cite{smartSSD9141369,reis_10.1145/3695053.3731116,CIPHERMATCH_10.1145/3676641.3716251,megis_10609570,genstore_10.1145/3503222.3507702,bluedbm_10.1145/2872887.2750412,mars_10.1145/3721145.3730428}.
Driven by this media-wide adaptability, researchers have proposed numerous application-specific PIM microarchitectures tailored for diverse workloads \cite{shared_pim_10979967,rowclone_10.1145/2540708.2540725,HBM-PIM-HW9365862,axdimm10.1145/3533737.3535093,megis_10609570,Polynesia9835628,mimdram_10476478,Functionally-Complete_10476435,sisa_10.1145/3466752.3480133,GraphPIM7920847}.
}


\revision{
In DRAM-based PIM systems, the absence of direct inter-PIM-unit interconnects forces data transfers to be mediated by CPU, incurring prohibitive communication overheads.
To bridge this gap, some works propose to introduce additional connections or custom access modes \cite{abndp10.1145/3582016.3582026,dimmlink10071005,ndp-bridge,abc-dimm9499805,PIMnet10946720}.
However, these methods introduce additional hardware complexity and require non-trivial adaptations at the DRAM interface.
Enabling PIM compatibility with commodity CPU ecosystems has also emerged as a major research focus.
To achieve this, subsequent works optimize system-level compatibility through software or protocol layers, such as deploying dynamic address mapping for shared memory spaces \cite{UM-PIM10609641,PIM-MMU10764703}, partitioning dedicated memory regions \cite{upmem8875680,HBM-PIM-HW9365862}, or implementing specialized cache coherence frameworks \cite{CoNDA8980366,MVC10.1145/3613424.3623784,LazyPIM7485993}.
These system-level protocols inevitably strain the memory control interface due to frequent host-PIM interactions.
To address this, frameworks like AsyncDIMM \cite{AsyncDIMM10946818} and ComPASS \cite{ComPASS10.1145/3725843.3756017} further alleviate host command issuance pressure and scheduling overheads.
Importantly, \archname{} operates orthogonally to these system and control interface optimizations, focusing on low-interference scheduling to further maximize overall system efficiency.
}

\revision{
\textbf{Memory Scheduling.} 
Conventional memory scheduling primarily focuses on mitigating row switching latencies \cite{moon2012compact,pret_6062323,salp_6237032} and read-write delays \cite{chen2012pre,Lee_TR_HPS_2010_007}. 
To scale within multi-core systems, recent works exploit individual thread variations to achieve fair and high-performance scheduling \cite{tcm5695526,fairscheduling4408252,FIQMR6332168,TBLMI7328201,parallelism-aware4556716,atlas5416658,fair_queuing_4041848,composable_5763145,lams7820660,BLISS_6974655,bliss2_7399423,firm_7011385,inter_wrap_gpu_7429292}, while also enforcing deadline-aware constraints \cite{bounding_6176554,real-time_10.1145/2435227.2435260,dash_10.1145/2847255}.
As architectures shift toward heterogenous systems (e.g., CPU-GPU), scheduling evolves to manage highly asymmetric core behaviors and requirements \cite{gpuheterosched10.1007/s11227-019-03135-7,inter-core7110523,clams10.1145/2896377.2901468,sms_10.5555/2337159.2337207,managing_gpu_concurrency_10.1109/MICRO.2014.62}.
To address unpredictable dynamic patterns, another research path deploys learning-based methods to adaptively optimize scheduling \cite{self-optimized4556714,rl-based10.1007/978-3-642-10684-2_10,epsilon_greedy_9478147}.
For PIM systems, the core difference is that CPU and PIM units do \emph{not} share a unified interface, operating via separate command sets that cause severe cross-domain interference and necessitate dedicated co-scheduling.
We have analyzed the two extreme policies in \sectionautorefname{}~\ref{sec:motivation:scheduling}, CPU-first \cite{Chopim9138972} and row-hit-aware \cite{AsyncDIMM10946818,F3FS11096393}, which biasedly favor CPU and PIM performance, respectively.
ComPASS \cite{ComPASS10.1145/3725843.3756017} introduces batch scheduling to balance performance, yet it enforces coarse-grained time-slicing between CPU and PIM, failing to achieve concurrent execution.
\archname{} simultaneously satisfies the low-latency requirements of CPU accesses while maximizing PIM throughput, minimizing interference during concurrent execution.
}

\section{Conclusion}
\label{sec:conclusion}

\revision{
    This paper introduces \archname{}, a cooperative memory access scheduling framework that accelerates LLM inference on mobile devices with marginal degradation to CPU performance. To achieve this, \archname{} tackles CPU-PIM resource contention from both the control interface and memory scheduling perspectives, dynamically leveraging host memory idle windows for PIM execution. Consequently, \archname{} achieves a 2.8\texttimes{} increase in LLM inference throughput with less than 2.0\% CPU performance degradation during concurrent execution.
}

\section*{Acknowledgments}

We sincerely thank the anonymous reviewers for their comments and suggestions to improve the paper.
This work was partially supported by the National Key Research and Development Program of China (2024YFE0204300), National Natural Science Foundation of China (Grant No.62402311), Natural Science Foundation of Shanghai (Grant No.24ZR1433700), and Key Research and Development Program of Shanghai (25LN3201200).
\appendix
\section{Artifact Appendix}

\subsection{Abstract}

This artifact provides the complete experimental framework for \archname{}, a cooperative scheduling framework designed for concurrent PIM/CPU execution on mobile devices. 
The artifact is built on the Ramulator2 simulator, which has been extended with the proposed low-interference PIM control interface and idleness-aware scheduling methodology. 

The software components include the modified C++ source code of the Ramulator2-based memory controller, specialized PIM instruction support, and the scheduling logic for PIM Execution Queues (PEQ) and PIM Read/Write Queues (PRWQ). 
To represent realistic mobile scenarios, we provide two types of memory traces: 
1) LLM workload traces (e.g., DeepSeek-R1-1.5B ) generated via Python scripts to simulate PIM tasks, and 2) Mobile application traces (e.g., Browser, YouTube ) collected from physical mobile devices to represent CPU tasks.

This artifact is designed to reproduce the key results presented in the paper, specifically the performance improvements and CPU interference analysis shown in Figures 8 and 10. 
Users can expect to observe that \archname{} significantly enhances PIM throughput (up to 2.8\texttimes ) while maintaining minimal CPU performance loss (less than 5.0\% ). 
The artifact requires a GCC 12+ compiler for building the simulator environment.

\subsection{Artifact check-list (meta-information)}


{\small
\begin{itemize}
  \item {\bf Algorithm: } Idleness-aware PIM scheduling, low-interference PIM control interface, and PIM/CPU concurrent memory access management.
  \item {\bf Program: } Modified Ramulator2 simulator (C++), trace generation scripts (Python).
  \item {\bf Compilation: } GCC 12 or higher (supporting C++20), CMake (3.30+).
  \item {\bf Data set: } LLM execution traces (DeepSeek-R1 series) and real-world mobile application traces (e.g., Browser, YouTube, Camera) collected from mobile devices.
  \item {\bf Run-time environment: } Linux (Ubuntu 20.04/22.04 recommended), Python 3.12+.
  \item {\bf Hardware: } Any x86\_64 CPU for simulation; approximately 64GB RAM recommended for large trace handling.
  \item {\bf Execution: } Automated simulation via Python task scripts; manual data integration into Excel templates.
  \item {\bf Metrics: } PIM throughput (Tokens/sec), CPU performance slowdown.
  \item {\bf Output: } Simulation log files, CSV tables, and generated plots (matching Figures 8, 10).
  \item {\bf Experiments: } Baseline vs. \archname{} comparison under different mobile scenarios; sensitivity analysis of PEQ/PRWQ sizes and scheduling thresholds.
  \item {\bf How much disk space required (approximately)?: } 5-10 GB (mostly for memory traces).
  \item {\bf How much time is needed to prepare workflow (approximately)?: } 30-60 minutes (for environment setup and simulator compilation).
  \item {\bf How much time is needed to complete experiments (approximately)?: } 1-2 hours.
  \item {\bf Publicly available?: } Yes (via Zenodo).
  \item {\bf Workflow automation framework used?: } Bash scripts.
  \item {\bf Archived (provide DOI)?: } \url{https://doi.org/10.5281/zenodo.19660293}
\end{itemize}
}

\subsection{Description}

\subsubsection{How to access}

The artifact is available via Zenodo. The repository includes the modified Ramulator2 source code, Python execution scripts, sample traces, and the Excel visualization template.


\subsubsection{Hardware dependencies}

The artifacts are based on software simulation. To reproduce the results within a reasonable timeframe (as specified in the check-list), a multi-core x86\_64 CPU (8+ cores) and at least 64GB of RAM are recommended.

\subsubsection{Software dependencies}

The simulator requires GCC 12 or higher to support C++20 standards. The experiment orchestration and data collection rely on Python 3.12+. For data visualization, we provide an Excel template that processes the generated CSV files.

\subsubsection{Data sets}

We include a comprehensive suite of memory traces:
\begin{itemize}
\item \textbf{CPU Workloads:} Real-world mobile application traces (e.g., YouTube, Browser) captured from a Xiaomi 11 Pro \cite{xiaomi_mi11pro_specs}.
\item \textbf{PIM Workloads:} LLM inference traces (e.g., DeepSeek-R1-1.5B) generated to simulate PIM-side memory access patterns.
\end{itemize}

\subsubsection{Models}

Our model extends Ramulator2 with the following \archname{}-specific features:
\begin{itemize}
\item Memory Controller: Implements the Idleness-aware scheduling logic, specialized queues (PEQ/PRWQ), and PIM/CPU command arbitration. 
(\path|src/dram_controller/impl/plugin/pim/pim_arbiter_cpu_first_o3_pred.cpp|)
\item PIM Control Interface: Adds support for low-interference PIM control commands.
(\path|src/dram/impl/LPDDR5-PIM.cpp|)
\end{itemize}

\subsection{Installation}


The installation of \archname{} follows the standard build procedure of Ramulator2.

1. Environment Setup:
The simulator requires a C++20 compliant compiler (GCC 12+) and CMake 3.30+. Detailed dependency requirements can be found in the official Ramulator2 documentation: \url{https://github.com/cmu-safari/ramulator2}. Alternatively, you can use the automated script provided in the repository:
\begin{lstlisting}[language=bash]
./build.sh
\end{lstlisting}
You can specify the parallelism level (default: 8) by running:
\begin{lstlisting}[language=bash]
./build.sh [parallelism]
# e.g. ./build.sh 16
\end{lstlisting}

2. Verification:
To ensure the build is successful, verify that the \texttt{ramulator2} executable is generated in the project root or the \path|build/| directory. You can run \path|./build/ramulator2 --help| to check the basic functionality.


\subsection{Experiment workflow}

The evaluation is conducted in three main stages: simulation execution, data aggregation, and visualization.

1. Simulation Execution:
We use a bash script \path|run_script.sh| to automate the simulation execution.
The script automatically manages the Ramulator2 execution and reads the PIM/CPU traces.
\begin{lstlisting}[language=bash]
cd simulations/
./run_script.sh figure8
./run_script.sh figure10
\end{lstlisting}
Alternatively, you can specify the parallelism level (default: 8) by running:
\begin{lstlisting}[language=bash]
./run_script.sh figure8 [parallelism]
# e.g. ./run_script.sh figure8 16
./run_script.sh figure10 [parallelism]
\end{lstlisting}
This will generate a CSV file under the \path|simulations/scripts_result_processor| directory.

2. Visualization:
Due to the complexity of the micro-architectural data analysis, we provide three Excel templates (\path:plot_figure8|10_ae.xlsx:) under \path|simulations/scripts_result_processor| directory for final plotting.
\begin{itemize}
\item Open the CSV file generated by the simulation (\path:summary\_figure8|10.csv:) and the corresponding Excel template.
\item Copy the content of the generated CSV file and paste it into the table \textbf{Summary}.
\item The corresponding charts in table \textbf{Chart} will automatically update to reflect the results, reproducing the trends shown in the paper.
\end{itemize}

\subsection{Evaluation and expected results}

The evaluation focuses on comparing the performance of \archname{} against baseline scheduling policies under concurrent PIM/CPU execution scenarios.
The reproduced results are expected to closely match the trends presented in the paper. \archname{} should consistently demonstrate higher PIM throughput (up to 2.8\texttimes) and lower CPU interference (within 5.0\% CPU performance loss).









\bibliographystyle{IEEEtran}
\bibliography{refs}

\end{document}